\documentclass[ aip, jmp, reprint,]{revtex4}
\usepackage{amsfonts}
\usepackage{url}
\usepackage{amssymb}
\usepackage{amsmath}
\usepackage{graphicx}
\usepackage{dcolumn}
\usepackage{bm}

\begin{document}

\title[]{\textbf{Hamiltonian formulation for the classical EM
radiation-reaction problem: \\ application to the kinetic theory
for relativistic collisionless plasmas.}}
\author{Claudio Cremaschini}
\affiliation{International School for Advanced Studies (SISSA), Trieste, Italy}
\affiliation{Consortium for Magnetofluid Dynamics, University of Trieste, Trieste, Italy}
\author{Massimo Tessarotto}
\affiliation{Department of Mathematics and Informatics, University of Trieste, Trieste,
Italy}
\affiliation{Consortium for Magnetofluid Dynamics, University of Trieste, Trieste, Italy}
\date{\today }

\begin{abstract}
A notorious difficulty in the covariant dynamics of classical charged
particles subject to non-local electromagnetic (EM) interactions arising in
the EM\ radiation-reaction (RR) phenomena is due to the definition of the
related \emph{non-local Lagrangian and Hamiltonian systems}. As a basic
consequence, the lack of a standard Lagrangian/Hamiltonian formulation in
the customary asymptotic approximation for the RR equation may inhibit the
construction of consistent kinetic and fluid theories. In this paper the
issue is investigated in the framework of Special Relativity. It is shown
that, for finite-size spherically-symmetric classical charged particles,
non-perturbative Lagrangian and Hamiltonian formulations in standard form
can be obtained, which describe particle dynamics in the presence of the
exact EM RR self-force. As a remarkable consequence, based on axiomatic
formulation of classical statistical mechanics, the covariant kinetic theory
for systems of charged particles subject to the EM RR self-force is
formulated in Hamiltonian form. A fundamental feature is that the non-local
effects enter the kinetic equation only through the retarded particle
4-position. This permits, in turn, the construction of the related fluid
equations, in which the non-local contributions carried by the RR effects
are explicitly displayed. In particular, it is shown that the moment
equations obtained in this way do not contain higher-order moments, allowing
as a consequence the adoption of standard closure conditions. A remarkable
aspect of the theory is related to the short delay-time asymptotic
expansions. Here it is shown that two possible expansions are permitted.
Both can be implemented for the single-particle dynamics as well as for the
corresponding kinetic and fluid treatments. In the last case, they are
performed \textit{a posteriori},\ namely on the relevant moment equations
obtained after integration of the kinetic equation over the velocity space.
Comparisons with the literature are pointed out.
\end{abstract}

\maketitle


\section{Introduction}

An open problem in relativistic theories is related to the Hamiltonian
description of particle dynamics for which non-local interactions typically
occur. In this regard, a basic difficulty which is usually met is the lack
of a Hamiltonian formalism for non-local Lagrangian systems. In fact, for
arbitrary non-local Lagrangians it is generally impossible to define the
notion of Legendre transformation \cite{Feytesi}. As a consequence even the
phase-space itself may not be well-defined.

Most approaches to the construction of a Hamiltonian formalism for non-local
first-order Lagrangians have tried to change the functional part of the
Euler-Lagrange equations \cite{Kerner1962,Marnelius1974,Gaida1980,Llosa1993}%
. In principle this delivers infinite-order Euler-Lagrange equations and a
corresponding infinite-dimensional phase-space. As an alternative, a finite
dimensional phase-space can be recovered by introducing appropriate
asymptotic approximations, i.e., truncating the expansion of the Lagrangian
in terms of finite-order derivatives \cite{Marnelius1974,LLosa1986}.

A typical situation of this kind occurs for the relativistic equation of
motion for single isolated charged particles, subject both to external and
self EM forces, namely the radiation-reaction (RR) equation. There is an
extensive literature devoted to this subject, most of which dealing with
point charges. As remarked by Dorigo \textit{et al.} \cite{Dorigo2008a},
customary formulations based either on the LAD \cite%
{Lorentz,Abraham1905,Dirac1938} or LL \cite{LL} equations are \emph{%
asymptotic,} i.e., obtained by means of asymptotic expansions of different
sort. In particular, as a consequence it follows that the LAD equation is
represented by a third-order ODE, so that it does not admit a Hamiltonian
formulation in the customary sense \cite{Goldstein,Arnold1976}. The LL,
instead, is \emph{intrinsically} non-variational, although it is a
second-order differential equation, being obtained by means of a one-step
\textquotedblleft reduction process\textquotedblright\ from the LAD equation
\cite{Dorigo2008a}. As a consequence, the LAD equation does not define a
dynamical system in the customary sense, since it requires, for non-rotating
particles, a 12-dimensional phase-space involving also the particle
acceleration. Therefore, for different reasons, both the LAD and LL
equations are \textit{manifestly non-Hamiltonian}. In particular, for the LL
equation, this implies that the corresponding phase-space volume is not
conserved. Moreover, within these treatments particles are treated as
point-like, so that non-local EM effects produced by the RR self-interaction
may remain undetermined.

Fundamental problems arise when attempting to formulate classical
statistical mechanics (CSM) for systems of relativistic charged particles
based on the LAD or LL equations. In fact even the proper axiomatic
formulation of the relativistic CSM for radiating particles is missing. This
requires the precise identification of the corresponding phase-space and the
definition of an invariant probability measure on this set. For a system of
charged particles subject solely to an external EM field and the RR
self-force this involves the construction of a Vlasov kinetic treatment. In
this regard, important issues concern:

1) The lack of a standard Hamiltonian formulation of relativistic CSM based
on such asymptotic equations, which implies the lack of a flow-preserving
measure. This feature is shared by both the LAD and LL equations.

2) The proper definition of a phase-space. The problem is relevant for the
LAD equation. In fact, although the construction of kinetic theory is still
formally possible \cite{Hakim,Hakim2}, the corresponding fluid statistical
description seems inhibited.

3) The explicit dependence of the kinetic distribution function (KDF) in
terms of the retarded EM self 4-potential are excluded. In fact, in the LAD
and LL approximations the self-potential does not appear explicitly (see for
example Refs.\cite{Tamburini,Ma2004}). Indeed, within the point-particle
model, underlying both treatments, the retarded self-potential is divergent.

On the other hand, for the fluid treatment:

1) The precise form of the fluid closure conditions may depend on the
approximations adopted in the kinetic description for the representation of
the EM RR self-force. An example-case is provided by Ref.\cite{Ma2004} where
relativistic fluid equations are obtained based on the LL equation. As a
result, it was found that, with the exception of the continuity equation,
all moment equations involve higher-order fluid moments associated to the RR
self-force. It is unclear whether this is an intrinsic physical feature of
the RR interaction or simply a result of the approximations involved.

2) The fluid fields may in principle depend implicitly on the EM self
4-potential. In the framework of the LL equation it is unclear how such an
effect can be dealt with. However, the treatment of such effects seems to
present objective difficulties. In fact, in principle non-local effects
might arise in this way in which \textit{retarded velocity contributions }%
appear in the kinetic equation. In such a case the explicit construction of
fluid equations would be ambiguous (and might involve an infinite set of
higher-order moments).

The interesting question is whether \emph{these difficulties can be overcome
in physically realizable situations, namely for exactly solvable classical
systems} (of particles)\emph{\ }for which\ the relativistic equations of
motion are both \emph{variational} and \emph{non-asymptotic}. The
prerequisite is provided by the possibility of constructing an exact
representation for the RR equation for a suitable type of classical charged
particles. In the past, their precise identification with
physically-realizable systems has remained elusive because of the difficulty
of the problem. However, as recently pointed out (see Tessarotto \textit{et
al.} \cite{Tessarotto2008a} and Cremaschini \textit{et al.} \cite%
{Cremaschini2011}, hereafter referred to as Paper I) in the framework of
special relativity an exact variational RR equation can be obtained for
classical finite-size charged particles. This refers to particles having a
finite-size mass and charge distributions which are quasi-rigid,
non-rotating, spherically symmetric and radially localized on a spherical
surface\emph{\ }$\partial \Omega $ having a finite radius $\sigma >0$ (see
\cite{Nodvik1964} and the related discussion in Paper I). In this
formulation, contrary to the point-particle case, the retarded EM self
4-potential is well-defined, namely, it does not diverge, and can be
determined analytically. As shown in Paper I, it follows that the RR
equation is variational and the corresponding Hamilton variational
functional is symmetric with respect to the non-local contributions. The
latter are due to the retarded EM self interaction arising from the finite
spatial extension of the charge distribution. As a consequence, the
resulting exact RR equation is a second-order delay-type ODE which admits a
Lagrangian formulation in standard form (see discussion below). Furthermore,
under suitable conditions, the same equation defines a classical dynamical
system (\textit{RR dynamical system}).

In this paper we intend to prove that, based on the results of Paper I, the
RR dynamical system admits also a Hamiltonian representation in terms of an
effective non-local Hamiltonian function $H^{eff}$. This implies that the
exact RR equation can also be cast in the equivalent\ \emph{standard
Hamiltonian form} represented by first-order delay-type ODEs%
\begin{eqnarray}
\frac{dr^{\mu }}{ds} &=&\frac{\partial H^{eff}}{\partial P_{\mu }},
\label{HAMILTON EQUATIONS} \\
\frac{dP_{\mu }}{ds} &=&-\frac{\partial H^{eff}}{\partial r_{\mu }},  \notag
\end{eqnarray}%
with $\mathbf{y}=\left( r^{\mu },P_{\mu }\right) $ denoting, in
superabundant variables, the particle canonical state which spans the
eight-dimensional phase-space $\Gamma \equiv \Gamma _{r}\times \Gamma _{u}$,
where $\Gamma _{r}$ and $\Gamma _{u}$ are respectively the Minkowski $M^{4}$%
-configuration space and the 4-dimensional velocity-space, both with metric $%
\eta _{\mu \nu }\equiv diag\left( 1,-1,-1,-1\right) $. Remarkably, here it
is found that the Hamiltonian structure can be retained also after the
introduction of a suitable short delay-time approximation of the RR force.
The result is an intrinsic feature of the extended particle model adopted in
the present treatment.

As a consequence, the statistical description of the RR dynamical system
follows in a standard way. In particular, here we report both the exact and
asymptotic kinetic and fluid formulations. These are developed for
collisionless relativistic plasmas in the Vlasov-Maxwell description,
including consistently the contribution carried by the RR self-field.
Applications of the theory here developed concern:

1) The kinetic and fluid treatments of relativistic astrophysical plasmas
observed, for example, in accretion disks, relativistic jets, active
galactic nuclei and mass inflows around compact objects.

2) The kinetic and fluid treatments of laboratory plasmas subject to
ultra-intense and pulsed-laser sources.

\subsection{Goals and scheme of the presentation}

In detail, the plan of the paper is as follows. In Section 2 we introduce
the basic definitions concerning the Lagrangian formulation for non-local
interactions and the concept of effective non-local Lagrangian function and
the related covariance properties (THM.1 and Corollary). In Section 3 we
provide an analogous generalization which permits to introduce the notion of
non-local Hamiltonian formulation (THM.2 and Corollary). In particular,
within the framework of the theory here developed, the standard form for the
local Legendre transformation is retained, while the concepts of effective
canonical momenta and effective Hamiltonian function are introduced. Then,
in Section 4 the general case of a rotating finite-size and
spherically-symmetric charged particle is discussed as a physical example of
non-local interaction. It is shown that the corresponding Hamilton
variational functional satisfies the requirements of THMs.1 and 2 (see
THM.3) and therefore admits both Lagrangian and Hamiltonian formulations in
standard form. In Section 5 the theory is applied to the specific case of a
non-rotating particle. As a result, based on the analytical results of Paper
I, the corresponding variational and effective Hamiltonian formulations are
presented (THM.4). This provides the explicit form of the effective
Hamiltonian function and a parameter-free representation for the retarded EM
RR self-force. Then, in Section 6 a Hamiltonian asymptotic approximation of
the RR equation is developed (THM.5), based on the retarded-time expansion
holding in the short delay-time ordering. The approximation overcomes basic
inconsistencies of the LAD and LL treatments, applying in the case of
point-particles. As a consequence, in Section 7 the relativistic kinetic
theory for a collisionless plasma with the inclusion of the EM RR effect is
formulated in Hamiltonian form. The use of superabundant canonical variables
allows a precise identification of the phase-space and the consequent
axiomatic formulation of CSM with non-local EM RR interactions. This is
based on the notion of invariant probability measure in such a setting. As a
result, a relativistic Liouville-Vlasov kinetic equation is proved to hold
for the KDF (THM.6). This permits to achieve a Vlasov-Maxwell description
applicable to relativistic plasmas, in which the RR interaction is
consistently taken into account also in the Maxwell equations both for the
external and self EM fields. In Section 8 the corresponding fluid fields and
fluid moment equations are computed in terms of 4-velocity integrals, which
retain the standard conservative Eulerian form as in the absence of RR
effects. The existence of both explicit and implicit non-local contributions
arising from the RR effect in the fluid equations is discussed. It is shown
that the former are associated to the EM force acting on the fluid, while
the latter enter in the definition of the fluid fields through the effective
momenta. In particular, the explicit dependence of the KDF on the retarded
EM self 4-potential is discussed and an asymptotic estimation of the
implicit contributions is presented. In Section 9 a Lagrangian formulation
of the fluid equations is derived, which allows one to introduce an explicit
parametrization of the non-local RR terms carried by the EM self 4-potential
and the EM RR force. It is shown that the exact fluid equations with the
inclusion of the RR interaction are of delay-type. Section 10 deals instead
with the development of asymptotic approximations of the moment equations.
This is motivated by the requirement of reducing the exact non-local fluid
equations to a local form. Different asymptotic approximations are obtained
for the non-local terms of the RR effect, based both on short-delay time
expansions (THM.7) and an iterative procedure which holds under the
assumption of weak RR self-force (in comparison with external EM and
pressure forces). A detailed analysis of the basic physical properties of
the kinetic and fluid treatments obtained here and a comparison with
previous literature is reported in Section 11. Concluding remarks are
presented in Section 12. Finally, in Appendix A a Green-function approach is
developed for the calculation of the EM self 4-potential, while in Appendix
B the connection with non-canonical representations is provided for the
relativistic kinetic theory.

\section{Non-local Lagrangian formulation}

The natural mathematical apparatus for an abstract description of Lagrangian
and Hamiltonian mechanics is that of variational principles, whose methods
have been studied for a long time by mathematicians and can be found in the
textbooks. Nevertheless, actual problems of interest in classical
relativistic dynamics involving the treatment of non-local interactions have
escaped a solution. In particular, in the literature the prevailing view is
that, while a non-local variational formulation is possible, a corresponding
Hamiltonian representation is generally excluded. In the following we intend
to point out that for a particular class of non-local Lagrangian systems the
problem can be given a complete solution. The latter correspond to
variational problems in which the variational functional is symmetric. To
this end, in this section we briefly recall basic notions holding for local
and non-local Lagrangian systems. This task represents a necessary
prerequisite for the establishment of a corresponding Hamiltonian
formulation and for the subsequent investigation of the Hamiltonian dynamics
of finite-size charged particles with the inclusion of the RR self-force.

\bigskip

\textbf{Definition \#1 - Local and non-local Lagrangian systems.}

A local (respectively, non-local) Lagrangian system is defined by the set $%
\left\{ \mathbf{x},L\right\} $ such that the following conditions are
satisfied.

\begin{enumerate}
\item $\mathbf{x\equiv }\left( r^{\mu }\left( s\right) ,\frac{dr^{\mu
}\left( s\right) }{ds}\equiv \overset{\cdot }{r}^{\mu }\right) $ is the
Lagrangian state spanning the Lagrangian phase space $\Gamma _{L}\subseteq
\mathbb{R}
^{2N}$.

\item The Lagrangian action functional $S$ is a 4-scalar of the form%
\begin{equation}
S=\int_{s_{1}}^{s_{2}}dsL,  \label{J-lagr}
\end{equation}%
with $L$ to be referred to as \emph{variational Lagrangian function}. In
particular, the functional dependencies of $S$ and $L$ are respectively of
the form:

\begin{itemize}
\item $S\equiv S_{0}\left( r\right) $ and $L\equiv L_{0}\left( r,\frac{dr}{ds%
}\right) $ for local systems;

\item $S\equiv S_{1}\left( r,\left[ r\right] \right) $ and $L\equiv
L_{1}\left( r,\frac{dr}{ds},\left[ r\right] \right) $ for non-local systems,
with $\left[ r\right] $ denoting non-local dependencies.
\end{itemize}

\item In the functional class
\begin{eqnarray}
\left\{ r^{\mu }\right\} &\equiv &\left\{ r^{\mu }(s):r^{\mu
}(s_{i})=r_{i}^{\mu },\text{ for }i=1,2,\text{ }s_{1},s_{2}\in I,\text{ }%
\right.  \notag \\
&&\left. \text{with }s_{1}<s_{2}\text{ and }r^{\mu }(s)\in C^{2}(I)\right\} ,
\label{fc}
\end{eqnarray}%
the synchronous variations $\delta r^{\mu }(s)$ are considered independent
and vanish at the endpoints $r^{\mu }(s_{i})=r_{i}^{\mu }$. Hereafter $%
\delta $ denotes, as usual, the Frechet functional derivative. For a
synchronous variational principle the interval $ds$ is such that $\delta
ds=0 $ and is subject to the constraint%
\begin{equation}
ds^{2}=g_{\mu \nu }dr^{\mu }\left( s\right) dr^{\nu }\left( s\right) ,
\label{cnstra}
\end{equation}%
where $r^{\mu }\left( s\right) $ are the extremal curves.

\item The Lagrangian action (\ref{J-lagr}) admits a unique extremal curve $%
r^{\mu }(s)$ such that, for all synchronous variations $\delta r^{\mu }(s)$
in the functional class (\ref{fc}) the Hamilton variational principle%
\begin{equation}
\delta S=0  \label{Hamiltonvariational principle II}
\end{equation}%
holds identically. For non-local systems the non-local Lagrangian must be
suitably constructed in such a way that the extrema curves $r^{\mu }\left(
s\right) $ satisfy the constraint (\ref{cnstra}).

In particular, for local systems the extremal curves of $S_{0}$ are provided
by the Euler-Lagrange (E-L) equations%
\begin{equation}
\frac{\delta S_{0}}{\delta r^{\mu }}\equiv F_{\mu }(r)L_{0}=0,
\label{standard part}
\end{equation}%
where, for an arbitrary set of Lagrange coordinates $q^{\mu },$ $F_{\mu }(q)$
denotes the \emph{E-L differential operator}%
\begin{equation}
F_{\mu }(q)\equiv \frac{d}{ds}\frac{\partial }{\partial \overset{\bullet }{q}%
^{\mu }}-\frac{\partial }{\partial q^{\mu }},  \label{E-L operator}
\end{equation}%
and $\overset{\bullet }{q}^{\mu }\equiv \frac{d}{ds}q^{\mu }\left( s\right)
. $

On the other hand, for non-local systems the extremal curves of the
functional $S_{1}$ are provided by the Euler-Lagrange equations%
\begin{equation}
\frac{\delta S_{1}}{\delta r^{\mu }}\equiv \left. \frac{\delta S_{1}}{\delta
r^{\mu }}\right\vert _{\left[ r\right] }+\left. \frac{\delta S_{1}}{\delta %
\left[ r^{\mu }\right] }\right\vert _{r}=0,  \label{varderlag}
\end{equation}%
where $\left. \frac{\delta S_{1}}{\delta r^{\mu }}\right\vert _{\left[ r%
\right] }$ and $\left. \frac{\delta S_{1}}{\delta \left[ r^{\mu }\right] }%
\right\vert _{r}$ carry respectively the contributions due to the local and
non-local dependencies.
\end{enumerate}

\bigskip

\textbf{Definition \#2 - Non-local Lagrangian systems in standard form.}

A non-local Lagrangian system $\left\{ \mathbf{x},L_{1}\right\} $ will be
said to admit a \emph{standard form} if the variational derivative (\ref%
{varderlag}) yields the E-L equations in\emph{\ the standard form:}%
\begin{equation}
\frac{\delta S_{1}}{\delta r^{\mu }}+\frac{\delta S_{1}}{\delta \left[
r^{\mu }\right] }\equiv F_{\mu }(r)L_{eff}=0,  \label{STANDARD}
\end{equation}%
with
\begin{equation}
L_{eff}\equiv L_{eff}\left( r,\frac{dr}{ds},\left[ r\right] \right)
\label{efflag}
\end{equation}%
denoting a suitable \emph{effective non-local Lagrangian.}

\bigskip

On the base of these definitions, the following theorem holds.

\bigskip

\textbf{THM.1 - Non-local and Effective Lagrangian functions}

\emph{Given validity of the definitions \#1 and \#2, it follows that:}

\emph{T1}$_{1}$\emph{) The non-local Lagrangian }$L_{1}$\emph{\ and the
effective Lagrangian }$L_{eff}$\emph{\ are generally different, namely}%
\begin{equation}
L_{1}\neq L_{eff}.  \label{l1leff}
\end{equation}

\emph{T1}$_{2}$\emph{) If }$S_{1}\left( r,\left[ r\right] \right) $\emph{\
admits the general decomposition}%
\begin{equation}
S_{1}(r,\left[ r\right] )=S_{a}(r)+S_{b}(r,\left[ r\right] ),  \label{j1dec}
\end{equation}%
\emph{with }$S_{a}(r)\equiv $ $\int_{s_{1}}^{s_{2}}dsL_{a}\left( r,\frac{dr}{%
ds}\right) $ \emph{and }$S_{b}(r,\left[ r\right] )\equiv
\int_{s_{1}}^{s_{2}}dsL_{b}\left( r,\frac{dr}{ds},\left[ r\right] \right) $%
\emph{, and moreover }$S_{b}(r,\left[ r\right] )$\emph{\ defines a symmetric
functional such that}%
\begin{equation}
S_{b}(r,\left[ r\right] )=S_{b}(\left[ r\right] ,r),  \label{sym1}
\end{equation}%
\emph{then the effective Lagrangian }$L_{eff}$\emph{\ is related to the
variational non-local Lagrangian }$L_{1}\equiv L_{a}+L_{b}$\emph{\ as}%
\begin{equation}
L_{eff}=L_{a}+2L_{b}=L_{1}+L_{b}.  \label{leff-lab}
\end{equation}

\emph{Proof} - T1$_{1}$) The proof is an immediate consequence of Eqs.(\ref%
{varderlag}) and (\ref{STANDARD}). In fact, by definition the E-L
differential operator $F_{\mu }(r)$ is a local differential operator that is
required to preserve its form also for non-local systems. On the other hand,
the variational derivative (\ref{varderlag}) is different from (\ref%
{standard part}). Hence, in order to write the E-L equations associated to
the non-local function $L_{1}$ in standard form, a suitable effective
Lagrangian $L_{eff}$ must be introduced, which must differ from $L_{1}$ and
be expressed in such a way that the non-local dependencies contained in $%
L_{1}$ can be equivalently treated by means of $F_{\mu }(r)$.

T1$_{2}$) The proof follows by inspecting the general definition (\ref%
{varderlag}). In this case, in view of the symmetry property (\ref{sym1}),
it follows manifestly that%
\begin{equation}
\frac{\delta S_{1}}{\delta r^{\mu }}\equiv \left. \frac{\delta S_{1}}{\delta
r^{\mu }}\right\vert _{\left[ r\right] }+\left. \frac{\delta S_{1}}{\delta %
\left[ r^{\mu }\right] }\right\vert _{r}=\left. \frac{\delta S_{a}}{\delta
r^{\mu }}\right\vert _{\left[ r\right] }+2\left. \frac{\delta S_{b}}{\delta
r^{\mu }}\right\vert _{\left[ r\right] }=0.
\end{equation}%
Then, by comparing this relation with the definitions both of the E-L
differential operator (\ref{E-L operator}) and the standard form
representation of the E-L equations (\ref{STANDARD}), it follows that the
effective Lagrangian $L_{eff}$ takes necessarily the form given in Eq.(\ref%
{leff-lab}). This completes the proof of the statement.

\textbf{Q.E.D.}

\bigskip

A basic consequence of Definition \#2 and THM.1 concerns the covariance
property of the E-L equations (\ref{STANDARD}). The result is stated in the
following Corollary.

\bigskip

\textbf{Corollary 1 to THM.1 - Covariance of the E-L equations for arbitrary
point transformations.}

\emph{The Euler-Lagrange equations (\ref{STANDARD}) are covariant with
respect to arbitrary point transformations}%
\begin{equation}
r^{\mu }\rightarrow q^{\mu }(r)  \label{pointtr}
\end{equation}%
\emph{represented by a diffeomeorphism of class }$C^{k}$, \emph{with }$k\geq
2,$ \emph{which requires they are of the form}%
\begin{equation}
F_{\mu }(q)\widetilde{L}_{eff}=\frac{\partial r^{\nu }}{\partial q^{\mu }}%
F_{\nu }(r)L_{eff}=0,  \label{elcov}
\end{equation}%
\emph{with }$\widetilde{L}_{eff}$\emph{\ denoting}
\begin{equation}
\widetilde{L}_{eff}\left( q,\frac{dq}{ds},\left[ q\right] \right) \equiv
L_{eff}\left( r,\frac{dr}{ds},\left[ r\right] \right) .
\end{equation}%
\emph{As a consequence, Eq.(\ref{STANDARD}) satisfies also the covariance
property with respect to arbitrary infinitesimal Lorentz transformations
(Manifest Lorentz Covariance).}

\emph{Proof}\ - The Euler-Lagrange equations (\ref{STANDARD}) are by
definition covariant provided the variational Lagrangian $L_{1}\left( r,%
\frac{dr}{ds},\left[ r\right] \right) $ is a 4-scalar (as it is by
construction). Then, it is sufficient to represent the Lagrangian action in
terms of the Lagrangian coordinates $q^{\mu },$ yielding%
\begin{equation}
\widetilde{S_{1}}(q,\left[ q\right] )\equiv S_{1}(r,\left[ r\right] ),
\end{equation}%
with $\widetilde{S_{1}}(q,\left[ q\right] )$ denoting the transformed action
\begin{equation}
\widetilde{S_{1}}(q,\left[ q\right] )=\int_{s_{1}}^{s_{2}}ds\widetilde{L_{1}}%
\left( q,\frac{dq}{ds},\left[ q\right] \right)
\end{equation}%
and $\widetilde{L_{1}}$ denoting the transformed variational non-local
Lagrangian. Hence, the Hamilton variational principle $\delta \widetilde{%
S_{1}}(q,\left[ q\right] )=0$ yields precisely the E-L equations (\ref{elcov}%
). This proves the statement. The covariance property of\ Eqs.(\ref{STANDARD}%
) with respect to point transformations (\ref{pointtr}) includes, as
particular case, Lorentz transformations. Therefore, Eqs.(\ref{STANDARD})
are also \emph{Manifestly Lorentz Covariant }(MLC).

\textbf{Q.E.D.}

\bigskip

We notice the following remarkable features of this treatment:

1) In general, in absence of any kind of symmetry, a non-local Lagrangian
system does not admit a standard form representation in terms of the
effective Lagrangian $L_{eff}$ \cite{Feytesi}.

2) As shown in T1$_{2}$, the possibility of getting an explicit relationship
between $L_{1}$ and $L_{eff}$ is a consequence solely of the symmetry
property (\ref{sym1}) of the functional $S_{b}$. This also proves the
existence of $L_{eff}$ and, as a consequence, of the standard form
representation for non-local systems satisfying Eq.(\ref{sym1}).

3) The symmetry assumption (\ref{sym1}) can be effectively realized in
physical systems. As it will be shown below, this condition is satisfied by
the variational functional which describes the dynamics of finite-size
classical charged particles with the inclusion of the RR effects associated
to the interaction with the EM self-field.

\bigskip

\section{Non-local Hamiltonian formulation}

In this section we deal with the basic features concerning the Hamiltonian
formulation for non-local systems which admit a variational treatment in
terms of non-local Lagrangian functions. This requires the introduction of
the following preliminary definitions.

\bigskip

\textbf{Definition \#3 - Local and non-local Hamiltonian systems.}

A local (respectively, non-local) Lagrangian system $\left\{ \mathbf{x}%
,L\right\} $ is said to admit a local (non-local) Hamiltonian system $%
\left\{ \mathbf{y\equiv (}r^{\mu }\mathbf{,}p_{\mu }\mathbf{),}H\right\} $
provide the following conditions are satisfied.

\begin{enumerate}
\item The \emph{variational Hamiltonian }$H$ is defined as the Legendre
transformation of the local (non-local) variational Lagrangian $L$
\begin{equation}
H=p_{\mu }\frac{dr^{\mu }}{ds}-L,  \label{canh}
\end{equation}%
with%
\begin{equation}
p_{\mu }=\frac{\partial L}{\partial \frac{dr^{\mu }}{ds}}  \label{canmom}
\end{equation}%
being the corresponding canonical momentum, with corresponding action
functional $S_{H}\equiv \int_{s_{1}}^{s_{2}}ds\left[ p_{\mu }\frac{dr^{\mu }%
}{ds}-H\right] $.

\item It is assumed that $H$ is respectively of the form:

\begin{itemize}
\item $H\equiv H_{0}\left( r,p\right) $ for local systems;

\item $H\equiv H_{1}\left( r,p,\left[ r\right] \right) $ for non-local
systems, namely it is a local function of $\left( r,p\right) $ and a
functional of $\left[ r\right] $.
\end{itemize}

The corresponding \emph{Hamilton action functionals} are denoted
respectively as%
\begin{equation}
S_{H_{0}}(r,p)=\int_{s_{1}}^{s_{2}}ds\left[ p_{\mu }\frac{dr^{\mu }}{ds}%
-H_{0}\right]
\end{equation}%
for local systems, and as%
\begin{equation}
S_{H_{1}}(r,p,\left[ r\right] )=\int_{s_{1}}^{s_{2}}ds\left[ p_{\mu }\frac{%
dr^{\mu }}{ds}-H_{1}\right]
\end{equation}%
for non-local systems.

\item In the functional class%
\begin{eqnarray}
\left\{ \mathbf{y\equiv (}r^{\mu }\mathbf{,}p_{\mu }\mathbf{)}\right\}
&\equiv &\left\{ \mathbf{y}(s):\mathbf{y}(s_{i})=\mathbf{y}_{i},\text{ for }%
i=1,2,\text{ }s_{1},s_{2}\in I,\text{ }\right.  \notag \\
&&\left. \text{with }s_{1}<s_{2}\text{ and }\mathbf{y}(s)\in C^{2}(I)\right\}
\end{eqnarray}%
the synchronous variations $\left( \delta r^{\mu }(s),\delta p_{\mu
}(s)\right) $ are all considered independent and vanish at the endpoints $%
\mathbf{y}(s_{i})=\mathbf{y}_{i}$. By assumption, synchronous variations
imply that $\delta ds=0$, with the interval $ds$ satisfying the constraint%
\begin{equation}
ds^{2}=g_{\mu \nu }dr^{\mu }\left( s\right) dr^{\nu }\left( s\right) ,
\end{equation}%
where $r^{\mu }\left( s\right) $ are the extremal curves.

\item The \emph{modified Hamilton variational principle}%
\begin{equation}
\delta S_{H}=0
\end{equation}%
with variations $\left( \delta r^{\mu }(s),\delta p_{\mu }(s)\right) $ is
equivalent to the Hamilton principle (\ref{Hamiltonvariational principle II}%
), i.e., it yields the same extremal curves in the functional class $\left\{
\mathbf{y}\right\} $.

In particular, for local systems the extremal curves of $S_{H_{0}}$ can be
cast in the \emph{standard Hamiltonian form as first-order ODEs}%
\begin{equation}
\frac{\delta S_{H_{0}}}{\delta p_{\mu }}=\frac{dr^{\mu }}{ds}=\frac{\partial
H_{0}}{\partial p_{\mu }}=\left[ r^{\mu },H_{0}\right] ,
\end{equation}%
\begin{equation}
\frac{\delta S_{H_{0}}}{\delta r^{\mu }}=-\frac{dp_{\mu }}{ds}=\frac{%
\partial H_{0}}{\partial r^{\mu }}=\left[ p_{\mu },H_{0}\right] ,
\end{equation}%
where the customary Poisson bracket formalism has been used.

On the other hand, for non-local systems the extremal curves of the
functional $S_{H_{1}}$ are provided by the set of first-order ODEs%
\begin{equation}
\frac{\delta S_{H_{1}}}{\delta p_{\mu }}=0,  \label{h1}
\end{equation}%
\begin{equation}
\frac{\delta S_{H_{1}}}{\delta r^{\mu }}\equiv \left. \frac{\delta S_{H_{1}}%
}{\delta r^{\mu }}\right\vert _{\left[ r\right] }+\left. \frac{\delta
S_{H_{1}}}{\delta \left[ r^{\mu }\right] }\right\vert _{r}=0,  \label{h2}
\end{equation}%
where $\left. \frac{\delta S_{H_{1}}}{\delta r^{\mu }}\right\vert _{\left[ r%
\right] }$ and $\left. \frac{\delta S_{H_{1}}}{\delta \left[ r^{\mu }\right]
}\right\vert _{r}$ carry respectively the contributions due to the local and
non-local dependencies.
\end{enumerate}

\bigskip

\textbf{Definition \#4 - Non-local Hamiltonian systems in standard form.}

A non-local Hamiltonian system $\left\{ \mathbf{y},H_{1}\right\} $ will be
said to admit a \emph{standard form} if the extremal first-order ODEs (\ref%
{h1}) and (\ref{h2}) can be cast in the \emph{standard Hamiltonian form} in
terms of the \emph{effective canonical momentum }$P_{\mu }$\emph{\ and
Hamiltonian function }$H^{eff}$ as%
\begin{equation}
\frac{\delta S_{H_{1}}}{\delta p_{\mu }}=\frac{dr^{\mu }}{ds}=\frac{\partial
H_{eff}}{\partial P_{\mu }}=\left[ r^{\mu },H_{eff}\right] ,  \label{djh1}
\end{equation}%
\begin{equation}
\frac{\delta S_{H_{1}}}{\delta r^{\mu }}\equiv \left. \frac{\delta S_{H_{1}}%
}{\delta r^{\mu }}\right\vert _{\left[ r\right] }+\left. \frac{\delta
S_{H_{1}}}{\delta \left[ r^{\mu }\right] }\right\vert _{r}=-\frac{dP_{\mu }}{%
ds}=\frac{\partial H_{eff}}{\partial r^{\mu }}=\left[ P_{\mu },H_{eff}\right]
.  \label{djh1-1}
\end{equation}%
Here both $H_{eff}=H_{eff}\left( r,P,\left[ r\right] \right) $ and $P_{\mu }$
must be defined in terms of the effective Lagrangian function introduced in
Eq.(\ref{STANDARD}) respectively as%
\begin{equation}
H_{eff}\equiv P_{\mu }\frac{dr^{\mu }}{ds}-L_{eff}  \label{heff}
\end{equation}%
and%
\begin{equation}
P_{\mu }\equiv \frac{\partial L_{^{eff}}}{\partial \frac{dr^{\mu }}{ds}}.
\label{pleff}
\end{equation}%
From this definition it follows that, if the non-local Hamiltonian system $%
\left\{ \mathbf{y},H_{1}\right\} $ admits a standard form, then the Poisson
bracket representation holds for $H_{eff}$ and $P_{\mu }$.

\bigskip

The following theorem can be stated concerning the relationship between $%
H_{1}$ and $H_{eff}$.

\bigskip

\textbf{THM.2 - Non-local and Effective Hamiltonian functions}

\emph{Given validity of the definitions \#3 and \#4 and the results of
THM.1, if }$S_{H_{1}}\left( r,p,\left[ r\right] \right) $\emph{\ admits the
general decomposition}%
\begin{equation}
S_{H_{1}}\left( r,p,\left[ r\right] \right) =S_{H_{a}}(r,p)+S_{H_{b}}(r,p,%
\left[ r\right] ),
\end{equation}%
\emph{with}%
\begin{eqnarray}
S_{H_{a}}(r,p) &\equiv &\int_{s_{1}}^{s_{2}}ds\left[ p_{a\mu }\frac{dr^{\mu }%
}{ds}-H_{a}\left( r,p\right) \right] , \\
S_{H_{b}}(r,p,\left[ r\right] ) &\equiv &\int_{s_{1}}^{s_{2}}ds\left[
p_{b\mu }\frac{dr^{\mu }}{ds}-H_{b}\left( r,p,\left[ r\right] \right) \right]
,
\end{eqnarray}%
\emph{where the canonical momenta }$p_{a\mu }$ \emph{and }$p_{b\mu }$ \emph{%
are defined respectively as}%
\begin{eqnarray}
p_{a\mu } &\equiv &\frac{\partial L_{^{a}}}{\partial \frac{dr^{\mu }}{ds}},
\\
p_{b\mu } &\equiv &\frac{\partial L_{^{b}}}{\partial \frac{dr^{\mu }}{ds}},
\end{eqnarray}%
\emph{and moreover }$S_{H_{b}}(r,p,\left[ r\right] )$\emph{\ defines a
symmetric functional such that}%
\begin{equation}
S_{H_{b}}(r,p,\left[ r\right] )=S_{H_{b}}(\left[ r\right] ,p,r),
\label{sym2}
\end{equation}%
\emph{then the effective Hamiltonian }$H_{eff}$\emph{\ is related to the
variational non-local Hamiltonian }$H_{1}\equiv H_{a}+H_{b}$\emph{\ as}%
\begin{equation}
H_{eff}=H_{a}+2H_{b}=H_{1}+H_{b},  \label{heff-hab}
\end{equation}%
\emph{where, by definition}%
\begin{eqnarray}
H_{1} &\equiv &p_{\mu }\frac{dr^{\mu }}{ds}-L_{1},  \label{h1bis} \\
H_{a} &\equiv &p_{a\mu }\frac{dr^{\mu }}{ds}-L_{a},  \label{ha} \\
H_{b} &\equiv &p_{b\mu }\frac{dr^{\mu }}{ds}-L_{b}.  \label{hb}
\end{eqnarray}

\emph{Proof} - The proof follows from THM.1 and by invoking the general
definitions (\ref{djh1}) and (\ref{djh1-1}). In fact, in view of the
symmetry property (\ref{sym2}), it follows manifestly that%
\begin{equation}
\frac{\delta S_{H_{1}}}{\delta r^{\mu }}\equiv \left. \frac{\delta S_{H_{1}}%
}{\delta r^{\mu }}\right\vert _{\left[ r\right] }+\left. \frac{\delta
S_{H_{1}}}{\delta \left[ r^{\mu }\right] }\right\vert _{r}=\left. \frac{%
\delta S_{H_{a}}}{\delta r^{\mu }}\right\vert _{\left[ r\right] }+2\left.
\frac{\delta S_{H_{b}}}{\delta r^{\mu }}\right\vert _{\left[ r\right] }.
\end{equation}%
Then, by comparing this relation with the definitions (\ref{heff})-(\ref%
{pleff}) for the standard Hamiltonian form and using Eqs.(\ref{h1bis})-(\ref%
{hb}), from the analogous results in THM.1 which concerns the relationship
between $L_{1}$ and $L_{eff}$ in the symmetric case, Eq.(\ref{heff-hab}) is
readily obtained.

\textbf{Q.E.D.}

\bigskip

Finally, as a basic consequence of Definition \#4 and THM.2, the following
Corollary can be stated concerning the covariance property of the Hamilton
equations in standard form.

\bigskip

\textbf{Corollary 1 to THM.2 - Covariance of the Hamilton equations for
arbitrary point transformations.}

\emph{The Hamilton equations (\ref{djh1})-(\ref{djh1-1}) in standard form
are covariant with respect to arbitrary point transformations}%
\begin{equation}
r^{\mu }\rightarrow q^{\mu }(r)  \label{pointtr-bis}
\end{equation}%
\emph{represented by a diffeomeorphism of class }$C^{k}$ \emph{with }$k\geq
2,$ \emph{which requires they are of the form}%
\begin{eqnarray}
\frac{dq^{\mu }}{ds} &=&\frac{\partial \widetilde{H}_{eff}}{\partial
P_{\left( q\right) \mu }}=\left[ q^{\mu },\widetilde{H}_{eff}\right] ,
\label{cov-h-1} \\
\frac{dP_{\left( q\right) \mu }}{ds} &=&-\frac{\partial \widetilde{H}_{eff}}{%
\partial q^{\mu }}=\left[ P_{\left( q\right) \mu },\widetilde{H}_{eff}\right]
,  \label{cov-h-2}
\end{eqnarray}%
\emph{with }$\widetilde{H}_{eff}$\emph{\ denoting}%
\begin{equation}
\widetilde{H}_{eff}\left( q,P,\left[ q\right] \right) \equiv H_{eff}\left(
r,P,\left[ r\right] \right)  \label{heff-tilda}
\end{equation}%
\emph{and }$P_{\left( q\right) \mu }$ \emph{being the transformed canonical
momentum. As a consequence, Eqs.(\ref{cov-h-1}) and (\ref{cov-h-2}) satisfy
also the covariance property with respect to arbitrary infinitesimal Lorentz
transformations (Manifest Lorentz Covariance).}

\emph{Proof}\ - In fact, for an arbitrary point transformation of the type (%
\ref{pointtr-bis}), the corresponding transformation for the momenta $P_{\nu
}$ is%
\begin{equation}
P_{(q)\mu }=\frac{\partial q^{\nu }}{\partial r^{\mu }}P_{\nu },
\end{equation}%
which yields%
\begin{eqnarray}
\frac{\partial P_{(q)\nu }}{\partial P_{\mu }} &=&\frac{\partial q^{\mu }}{%
\partial r^{\nu }}, \\
\frac{\partial P_{(q)\mu }}{\partial P_{\nu }} &=&\frac{\partial r^{\nu }}{%
\partial q^{\mu }}.
\end{eqnarray}%
Hence, it follows that%
\begin{eqnarray}
\frac{dq^{\mu }}{ds} &=&\frac{\partial \widetilde{H}_{eff}}{\partial
P_{(q)\mu }}=\frac{\partial q^{\mu }}{\partial r^{\nu }}\frac{dr^{\nu }}{ds}=%
\frac{\partial q^{\mu }}{\partial r^{\nu }}\frac{\partial H_{eff}}{\partial
P_{\nu }}, \\
\frac{dP_{(q)\mu }}{ds} &=&-\frac{\partial \widetilde{H}_{eff}}{\partial
q^{\mu }}=\frac{\partial P_{(q)\mu }}{\partial P_{\nu }}\frac{dP_{\nu }}{ds}%
=-\frac{\partial r^{\nu }}{\partial q^{\mu }}\frac{\partial H_{eff}}{%
\partial r^{\nu }},
\end{eqnarray}%
which implies
\begin{eqnarray}
\frac{\partial \widetilde{H}_{eff}}{\partial P_{(q)\mu }} &=&\frac{\partial
q^{\mu }}{\partial r^{\nu }}\frac{\partial H_{eff}}{\partial P_{\nu }},
\label{htilde-cov1} \\
\frac{\partial \widetilde{H}_{eff}}{\partial q^{\mu }} &=&\frac{\partial
r^{\nu }}{\partial q^{\mu }}\frac{\partial H_{eff}}{\partial r^{\nu }},
\label{htilde-cov2}
\end{eqnarray}%
where $\widetilde{H}_{eff}$ is defined in Eq.(\ref{heff-tilda}) above.
Therefore, the Hamilton equations in standard form for the Lagrangian
coordinates $q^{\mu }$\ and the canonical momenta $P_{(q)\mu }$\ are
respectively covariant [Eq.(\ref{htilde-cov1})] and controvariant [Eq.(\ref%
{htilde-cov2})]\ with respect to the point transformation (\ref{pointtr-bis}%
). This is true also for arbitrary infinitesimal Lorentz transformations,
which proves the MLC of Hamilton equations Eqs.(\ref{cov-h-1}) and (\ref%
{cov-h-2}) in standard form.

\textbf{Q.E.D.}

\bigskip

\section{An example of non-local interaction: the classical EM RR problem}

A crucial issue of the present investigation concerns the possible existence
of physical systems subject to non-local interactions whose dynamics can be
consistently described in terms of a variational action integral and which
admit at the same time both Lagrangian and Hamiltonian formulations in
standard form. In this section we prove that the EM RR problem for classical
finite-size charged particles represents a physical example of non-local
interactions of this kind. The reason behind the choice of considering
extended particles is the necessity of avoiding the intrinsic divergences of
the RR effect characteristic of the point-particle model.

In fact, consider the general form of the Hamilton action functional for the
variational treatment of the dynamics of an extended charged particle in
presence of an external EM field and with the inclusion of the RR
self-interaction. This can be conveniently expressed as follows:
\begin{equation}
S_{1}\left( z,\left[ z\right] \right) =S_{M}\left( z\right)
+S_{C}^{(ext)}(z)+S_{C}^{(self)}(z,\left[ z\right] ),  \label{s1}
\end{equation}%
where $S_{M},S_{C}^{(ext)}$and $S_{C}^{(self)}$ are respectively the
contributions from the inertial mass and the EM coupling with the external
and the self fields. In particular, denoting by $j^{\left( self\right) \mu
}(r)$ the particle 4-current density generated by the particle itself and
observed at a 4-position $r$, the two coupling action integrals are provided
by the following 4-scalars:%
\begin{eqnarray}
S_{C}^{(ext)}\left( z\right) &=&\frac{1}{c^{2}}\int_{1}^{2}d^{4}rA^{(ext)\mu
}\left( r\right) j_{\mu }^{\left( self\right) }(r),  \label{sext} \\
S_{C}^{(self)}(z,\left[ z\right] ) &=&\frac{1}{c^{2}}%
\int_{1}^{2}d^{4}rA^{(self)\mu }\left( r\right) j_{\mu }^{\left( self\right)
}(r),  \label{sself}
\end{eqnarray}%
where $A_{\mu }^{(ext)}$ and $A_{\mu }^{(self)}$ denote the 4-vector
potentials of the external and the self EM fields and $z$ is a state to be
suitably defined (see below). A clarification here is in order. The external
EM 4-potential $A_{\mu }^{(ext)}\left( r\right) $ acting on the charged
particle located at the 4-position $r$ is assumed to be produced only by
prescribed \textquotedblleft external\textquotedblright\ sources, namely,
excluding the particle itself, by the remaining possible EM sources
belonging to the configuration space $\Gamma _{r}$. Within the framework of
special relativity, both the inertial term and the coupling term with the
external field carry only local dependencies, in the sense that they depend
explicitly only on the local 4-position $r$. They provide the classical
dynamics of charged particles in absence of any RR effect. On the other
hand, the functional $S_{C}^{(self)}$ associated to the EM self-interaction
contains both local and non-local contributions. In particular, since the
state $z$ of a finite-size particle must include a 4-position vector $r$, it
follows that $S_{C}^{(self)}$ generally depends explicitly on two different
4-positions, $r$ and $\left[ r\right] $, to be properly defined (see below).
The non-local property of $S_{C}^{(self)}$ represents a characteristic
feature of RR phenomena.

From the relationship (\ref{s1}) it follows that the Hamilton action
functional for the treatment of the RR admits the decomposition (\ref{j1dec}%
) introduced by THM.1, namely it can be written as the sum of two terms,
carrying respectively only local and both local and non-local dependencies.
In order to prove that the same functional admits also a Lagrangian and a
Hamiltonian representation in standard form it is sufficient to show that
the self-coupling functional is symmetric in $z$ and $\left[ z\right] $, in
the sense defined in THM.1. For this purpose we need to determine explicitly
the general expression of the 4-current and the self 4-potential for a
rotating finite-size charged particle.

The first step consists in constructing a covariant representation for the
4-current density. We follow the approach presented by Nodvik \cite%
{Nodvik1964}. Thus, we consider an extended charged particle with charge and
mass distributions having the same support $\partial \Omega $, to be
identified with a smooth surface. Denoting by $r^{\mu }\left( s\right) $ the
4-vector position (with proper time $s$) of a reference point belonging to
the internal open domain $\Omega $ and by $\zeta ^{\mu }$ a generic 4-vector
of $\partial \Omega $, the displacement vector $\xi ^{\mu }$ is defined as:%
\begin{equation}
\xi ^{\mu }\equiv \varsigma ^{\mu }-r^{\mu }(s).
\end{equation}%
The particle model is prescribed by imposing the constraints of rigidity of $%
\partial \Omega $, namely for all $\varsigma ^{\mu }$ and $r^{\mu }\left(
s\right) $ \cite{Nodvik1964}:%
\begin{eqnarray}
\xi ^{\mu }\xi _{\mu } &=&const.,  \label{eee1a} \\
\xi _{\mu }u^{\mu }(s) &=&0,  \label{eee2a}
\end{eqnarray}%
where $u^{\mu }(s)\equiv \frac{d}{ds}r^{\mu }(s)$. In particular, we shall
assume that mass and charge distributions are spherically symmetric and
therefore characterized by a form factor $f\left( \xi ^{2}\right) \equiv
f\left( \xi ^{\mu }\xi _{\mu }\right) $. This allows one to identify $r^{\mu
}\left( s\right) $ as the center-point of $\partial \Omega $. The extended
particle can in principle exhibit both translational and rotational degrees
of freedom. In particular, the translational motion can be described in
terms of $r^{\mu }\left( s\right) $. Instead, the rotational dynamics, which
includes both space-time rotations associated to the so-called Thomas
precession and pure spatial rotations, can be described in terms of the
Euler angles $\alpha (s)\equiv \left\{ \varphi (s),\vartheta (s),\psi
(s)\right\} $. It follows that, in this case, the Lagrangian state $z$ must
be identified with the set of variables $z\equiv \left( r^{\alpha }\left(
s\right) ,\alpha (s)\right) $. In view of these definitions it is immediate
to prove that the 4-current density for the finite-size particle can be
written as follows:%
\begin{equation}
j^{\left( self\right) \mu }(r)=qc\int_{-\infty }^{+\infty }ds\left\{ u^{\mu
} \left[ 1-\frac{du_{\alpha }}{ds}x^{\alpha }\right] -\frac{1}{c}\omega
^{\mu \nu }x_{\nu }\right\} f(x^{2})\delta (x^{\alpha }u_{\alpha }),
\label{jmu}
\end{equation}%
where%
\begin{equation}
x^{\mu }=r^{\mu }-r^{\mu }\left( s\right)  \label{xx}
\end{equation}%
and $\omega ^{\mu \nu }=\omega ^{\mu \nu }\left( s\right) $ is the
antisymmetric angular velocity tensor \cite{Nodvik1964}, which depends on $s$
through the Euler angles $\alpha (s)$. The term $\left[ 1+\frac{du_{\alpha }%
}{ds}x^{\alpha }\right] $ contains the acceleration of $r^{\mu }\left(
s\right) $ and represents the contribution associated to the Thomas
precession effect. This can be formally eliminated by using the properties
of the Dirac-delta function, implying the identity:%
\begin{equation}
\delta (x^{\alpha }u_{\alpha }(s))=\frac{1}{\left\vert \frac{d\left[
x^{\alpha }u_{\alpha }\right] }{ds}\right\vert }\delta (s-s_{1})=\frac{1}{%
\left\vert 1-\frac{du_{\alpha }}{ds}x^{\alpha }\right\vert }\delta (s-s_{1}),
\label{prop}
\end{equation}%
where by definition $s_{1}=s_{1}\left( r\right) $ is the root of the
algebraic equation%
\begin{equation}
u_{\mu }(s_{1})\left[ r^{\mu }-r^{\mu }\left( s_{1}\right) \right] =0.
\end{equation}%
As a result, the 4-current can be equivalently expressed as%
\begin{equation}
j^{\left( self\right) \mu }(r)=qc\int_{-\infty }^{+\infty }ds\left[ u^{\mu
}\delta (s-s_{1})-\frac{1}{c}\omega ^{\mu \nu }x_{\nu }\delta (x^{\alpha
}u_{\alpha })\right] f(x^{2}).  \label{jmu2}
\end{equation}

The second step consists in constructing a Green-function representation for
the EM self-potential $A^{(self)\mu }$ in terms of the 4-current $j^{\left(
self\right) \mu }(r)$. This technique is well-known. Thus, considering the
Maxwell equations in flat space-time, in the Lorentz gauge $A^{(self)\beta
},_{\beta }=0$, the self 4-potential must satisfy the wave equation%
\begin{equation}
\square A^{(self)\mu }=\frac{4\pi }{c}j^{\left( self\right) \mu }(r),
\label{6biz}
\end{equation}%
where $\square $ represents the D'Alembertian operator and $j^{\left(
self\right) \mu }(r)$ is given by Eq.(\ref{jmu2}). The formal solution of
Eq.(\ref{6biz}) is%
\begin{equation}
A^{(self)\mu }(r)=\frac{4\pi }{c}\int d^{4}r^{\prime }G(r,r^{\prime
})j^{\left( self\right) \mu }(r^{\prime }),  \label{D-1}
\end{equation}%
where $G(r,r^{\prime })$ is the retarded Green's function corresponding to
the prescribed charge density. By construction, it follows that $G\left(
r,r^{\prime }\right) $ is symmetric with respect to $r$ and $r^{\prime }$,
and furthermore - since the particle is finite-size - both the 4-current and
the self-potential are everywhere well-defined.

From these general results, it is immediate to prove the following theorem.

\bigskip

\textbf{THM.3 - Symmetry properties of }$S_{C}^{(self)}(z,\left[ z\right] )$

\emph{Given validity of Eq.(\ref{jmu2}) for the covariant expression of the
current density for a finite-size charged particle and of Eq.(\ref{D-1}) for
the general expression of the corresponding EM self-potential, it follows
that:}

\emph{T3}$_{1}$\emph{) The functional }$S_{C}^{(self)}(z,\left[ z\right] ),$
\emph{defined in Eq.(\ref{sself}) as an integral over the 4-volume element }$%
d^{4}r,$ \emph{can be written as a line integral of the form}%
\begin{equation}
S_{C}^{(self)}(z,\left[ z\right] )=\int_{-\infty }^{+\infty
}dsL_{C}^{(self)}\left( z,\left[ z\right] \right) ,  \label{scself-thm3}
\end{equation}

\emph{where }$L_{C}^{(self)}$\emph{\ represents the Lagrangian of the
coupling with the EM self-field. This is defined as}%
\begin{equation}
L_{C}^{(self)}\left( z,\left[ z\right] \right) \equiv \frac{4\pi q}{c^{2}}%
\int_{1}^{2}d^{4}r\left\{ \left[ u^{\mu }\delta (s-s_{1})-\frac{1}{c}\omega
^{\mu \nu }x_{\nu }\delta (x^{\alpha }u_{\alpha })\right] f(x^{2})\int
d^{4}r^{\prime }G(r,r^{\prime })j_{\mu }^{\left( self\right) }(r^{\prime
})\right\} .  \label{lcself-thm3}
\end{equation}

\emph{T3}$_{2}$\emph{) The functional }$S_{C}^{(self)}(z,\left[ z\right] )$
\emph{contains both local and non-local dependencies in terms of the
variational quantities }$z\equiv z\left( s\right) $\emph{\ and} $\left[ z%
\right] \equiv \left[ z\left( s\right) \right] .$\emph{\ In particular, it
is symmetric in these local and non-local variables, in the sense stated in
THM.1, namely}%
\begin{equation}
S_{C}^{(self)}(z,\left[ z\right] )=S_{C}^{(self)}(\left[ z\right] ,z).
\label{sym-thm3}
\end{equation}

\emph{T3}$_{3}$\emph{) The functional }$S_{C}^{(self)}(z,\left[ z\right] )$
\emph{contains at most only first-order derivatives of the variational
functions }$z\left( s\right) $.

\emph{Proof} - T3$_{1}$) The proof of the first statement follows by noting
that the action integral $S_{C}^{(self)}(z,\left[ z\right] )$ is a 4-scalar
by definition. Hence, making explicit the expressions of $A^{(self)\mu }$
and $j_{\mu }^{\left( self\right) }$ in Eq.(\ref{sself}) according to the
results in Eqs.(\ref{D-1}) and (\ref{jmu2}), by exchanging the order of the
integrations and invoking the symmetry property of the Green function, the
conclusion can be easily reached. In particular, the variational Lagrangian
is found to be of the general form given in Eq.(\ref{lcself-thm3}).

T3$_{2}$) To prove the second statement we first notice that in Eq.(\ref%
{scself-thm3}) both $z$ and $z^{\prime }$ are integration variables, while
by definition the variational quantities are identified with $z\left(
s\right) $\emph{\ }and $\left[ z\left( s\right) \right] \equiv z^{\prime
}\left( s^{\prime }\right) $. These dependencies are carried respectively by
the charge current densities $j^{\left( self\right) \mu }(r)$ and $j^{\left(
self\right) \mu }(r^{\prime })$. The result is then reached by noting that
the functional carrying the self-coupling terms is symmetric with respect to
the integrated quantities, and in particular with respect to $j^{\left(
self\right) \mu }(r)$ and $j^{\left( self\right) \mu }(r^{\prime })$. Hence,
exchanging $\left( z,j^{\left( self\right) \mu }(r)\right) $ with $\left(
z^{\prime },j^{\left( self\right) \mu }(r^{\prime })\right) $ does not
affect the form of the functional, with the consequence that Eq.(\ref%
{sym-thm3}) is identically satisfied.

T3$_{3}$) The proof of the statement is an immediate consequence of the
representation for the current density $j^{\left( self\right) \mu }(r)$
given in Eq.(\ref{jmu2}). In fact, the term proportional to the acceleration
$\frac{du_{\alpha }}{ds}$ in Eq.(\ref{jmu}) and which is associated to the
Thomas precession, does not appear in Eq.(\ref{jmu2}), thanks to the
property of the Dirac-delta function indicated above in Eq.(\ref{prop}).

\textbf{Q.E.D.}

\bigskip

An immediate consequence of THM.3 is that, thanks to THMs.1 and 2, the
variational treatment of the dynamics of finite-size charged particles
subject to the EM RR effect admits both Lagrangian and Hamiltonian
representations in standard form. In particular, in this case, it follows
that the following identification must be introduced:%
\begin{equation}
L_{b}\equiv L_{C}^{(self)},
\end{equation}%
where $L_{b}$ is the Lagrangian defined above in THM.1.

\bigskip

\section{Hamiltonian theory for the RR problem}

In this section, based on THMs.1-3 and the theory developed in Paper I, we
proceed constructing the Hamiltonian formulation for the RR problem. For
this purpose, it is convenient to recall the explicit form of the EM self
4-potential obtained in Paper I. For the sake of comparison with traditional
approaches based on point particle models, here we also propose an
alternative approach based on the Green function method. Remarkably, as
pointed out in Appendix A, for the spherically-symmetric and non-rotating
extended particle considered here, the self-potential is proved to be
formally analogous to the well-known solution valid for point charges. This
result holds, however, only in the external domain (with respect to $%
\partial \Omega $), where $A_{\mu }^{(self)}(r)$ is found to admit the
integral representation (see details in Appendix A):%
\begin{equation}
A_{\mu }^{(self)}(r)=2q\int_{1}^{2}dr_{\mu }^{\prime }\delta (\widehat{R}%
^{\alpha }\widehat{R}_{\alpha }).  \label{intrepA}
\end{equation}%
Here $\widehat{R}^{\alpha }=r^{\alpha }-r^{\alpha }(s^{\prime })$, with $%
r^{\alpha }$ and $r^{\prime \alpha }\equiv r^{\alpha }(s^{\prime })$
denoting respectively the generic 4-position and the 4-position of the
center of the charge distribution at proper time $s^{\prime }$. As a
fundamental consequence of the finite extension of the particle and the
restrictions on the domain of validity of Eq.(\ref{intrepA}), the resulting
variational functional and Faraday tensor for the self-field turn out to be
completely different from the point-particle treatment. In particular, the
action integral becomes now a non-local functional with respect to the
4-position $r$. As pointed out in Paper I, this can be written as a line
integral in terms of a variational Lagrangian $L_{1}(r,\left[ r\right] )$ as
follows:%
\begin{equation}
S_{1}(r,\left[ r\right] )=\int_{-\infty }^{+\infty }dsL_{1}(r,\left[ r\right]
).
\end{equation}%
Here $L_{1}(r,\left[ r\right] )$ is defined as:%
\begin{equation}
L_{1}(r,\left[ r\right] )=L_{M}(r)+L_{C}^{(ext)}(r)+L_{C}^{(self)}(r,\left[ r%
\right] ),  \label{EXTREMAL LAGRANGIAN}
\end{equation}%
where%
\begin{eqnarray}
L_{M}(r,u) &=&\frac{1}{2}m_{o}c\frac{dr_{\mu }}{ds}\frac{dr^{\mu }}{ds},
\label{LAGRANGIAN -constarint-2} \\
L_{C}^{(ext)}(r) &=&\frac{q}{c}\frac{dr}{ds}^{\mu }\overline{A}_{\mu
}^{(ext)}(r),  \label{LAGRANGIAN -EXTERNAL EM}
\end{eqnarray}%
are the local contributions respectively from the inertial and the external
EM field coupling terms, with $\overline{A}_{\mu }^{(ext)}$ denoting the
surface-averaged external EM potential (see Paper I for its definition). On
the other hand, $L_{C}^{(self)}$ represents the non-local contribution
arising from the EM self-field coupling, which is provided by%
\begin{equation}
L_{C}^{(self)}(r,\left[ r\right] )=\frac{2q^{2}}{c}\frac{dr}{ds}^{\mu
}\int_{1}^{2}dr_{\mu }^{\prime }\delta (\widetilde{R}^{\mu }\widetilde{R}%
_{\mu }-\sigma ^{2}),  \label{LAGRANGIAN-SELF-EM}
\end{equation}%
where the 4-scalar $\sigma ^{2}\equiv \xi ^{\mu }\xi _{\mu }$ is the radius
of the surface distribution with respect to the center $r^{\mu }\left(
s\right) $ and $\widetilde{R}^{\mu }$ is defined as%
\begin{equation}
\widetilde{R}^{\alpha }\equiv r^{\alpha }\left( s\right) -r^{\alpha
}(s^{\prime }).
\end{equation}%
Notice that $\widetilde{R}^{\alpha }$ represents the displacement bi-vector
between the actual position $r^{\alpha }\left( s\right) $ of the charge
center at proper time $s$ and the retarded position $r^{\alpha }(s^{\prime
}) $ of the same point at the retarded proper time $s^{\prime }$. It is
immediate to verify that the representation of $S_{C}^{(self)}$ in terms of $%
L_{C}^{(self)}$ given in Eq.(\ref{LAGRANGIAN-SELF-EM}) satisfies the
hypothesis of THM.1, and therefore the solution admits a Lagrangian
representation in standard form. As already shown in Paper I and according
to THM.1, this is obtained by setting%
\begin{equation}
L_{eff}\equiv L_{M}(r)+L_{C}^{(ext)}(r)+2L_{C}^{(self)}(r,\left[ r\right] ),
\label{extremal Lagrangian-0}
\end{equation}%
with $L_{M}(r)$, $L_{C}^{(ext)}(r)$ and $L_{C}^{(self)}$ respectively given
by Eqs.(\ref{LAGRANGIAN -constarint-2})-(\ref{LAGRANGIAN-SELF-EM}). Then,
the corresponding E-L equation is provided by the following covariant
4-vector, second-order delay-type ODE:
\begin{equation}
m_{o}c\frac{du_{\mu }(s)}{ds}=\frac{q}{c}\overline{F}_{\mu \nu
}^{(ext)}(r(s))\frac{dr^{\nu }(s)}{ds}+\frac{q}{c}\overline{F}_{\mu
k}^{\left( self\right) }\left( r\left( s\right) ,r\left( s^{\prime }\right)
\right) \frac{dr^{k}(s)}{ds},  \label{RR}
\end{equation}%
where%
\begin{equation}
u^{\mu }\left( s\right) \equiv \frac{dr^{\mu }(s)}{ds}.  \label{4vel}
\end{equation}%
Here the notation is as follows. Denoting by $F_{\mu \nu }\equiv F_{\mu \nu
}^{(ext)}+F_{\mu \nu }^{(self)}$ the total Faraday tensor, $F_{\mu \nu
}^{(ext)}$ and $F_{\mu \nu }^{(self)}$ are respectively the
\textquotedblleft external\textquotedblright\ and \textquotedblleft
self\textquotedblright\ Faraday tensors generated by $A_{\nu }^{(ext)}$ and $%
A_{\nu }^{(self)}$, which carry the contributions due to the external
sources with respect to the charged particle and the particle EM
self-interaction. In particular, the 4-tensor $\overline{F}_{\mu \nu
}^{(ext)}(r(s))$ denotes the surface-average of the Faraday tensor
associated to the external EM field, to be identified with%
\begin{equation}
\overline{F}_{\mu \nu }^{(ext)}\equiv \partial _{\mu }\overline{A}_{\nu
}^{(ext)}-\partial _{\nu }\overline{A}_{\mu }^{(ext)},
\label{EXTERNAL EM FIELD}
\end{equation}%
with $\overline{A}_{\nu }^{(ext)}(r(s))$ only generated by external sources
with respect to the single-particle whose dynamics is described by Eq.(\ref%
{RR}). Similarly, $\overline{F}_{\mu k}^{\left( self\right) }$\ is the
surface-average of the Faraday tensor contribution carried by the EM self
4-potential. In the parameter-free representation this is given by%
\begin{equation}
\overline{F}_{\mu k}^{\left( self\right) }(r,\left[ r\right]
)=-4q\int_{1}^{2}\left[ dr_{\mu }^{\prime }\frac{\partial }{\partial r^{k}}%
\delta \left( \widetilde{R}^{\alpha }\widetilde{R}_{\alpha }-\sigma
^{2}\right) -dr_{k}^{\prime }\frac{\partial }{\partial r^{\mu }}\delta
\left( \widetilde{R}^{\alpha }\widetilde{R}_{\alpha }-\sigma ^{2}\right) %
\right] .  \label{COVARIANT FORM}
\end{equation}%
As pointed out in Paper I, $\overline{F}_{\mu k}^{\left( self\right) }$ can
also be parametrized in terms of the particle proper time $s,$ by letting $%
r\equiv r\left( s\right) $ and $\left[ r\right] \equiv r\left( s^{\prime
}\right) $ in the previous equation, which also implies $dr_{\mu }^{\prime
}\equiv ds^{\prime }\frac{dr_{\mu }^{\prime }}{ds^{\prime }}$. This means
that the non-locality in Eq.(\ref{COVARIANT FORM}) can be interpreted as
non-locality in the particle proper time.

The remarkable feature of Eq.(\ref{COVARIANT FORM}) is that the RR
self-force (see the second term in the rhs of Eq.(\ref{RR})) contains
non-local effects only through the retarded particle 4-position and not
through the 4-velocity. This feature is fundamental for the subsequent fluid
treatment, since it permits the evaluation in the standard way of the
velocity moments, retaining the exact form of the RR self-interaction.

The system of Eqs.(\ref{RR}) and (\ref{4vel}) defines a delay-type ODE
problem of the form%
\begin{equation}
\left\{
\begin{array}{c}
\frac{d\mathbf{y}}{ds}=\mathbf{X}_{H}\left( \mathbf{y},\left[ r\right]
\right) , \\
\mathbf{y}\left( s_{0}\right) =\mathbf{y}_{0}, \\
\mathbf{y}\left( s_{0}^{\prime }\right) =\mathbf{y}_{s_{0}^{\prime }},\ \ \
\forall s_{0}^{\prime }\in I_{s_{0},s_{0}-s_{ret}},%
\end{array}%
\right.  \label{initial pro}
\end{equation}%
with $s_{0}$ and $s_{ret}$ denoting respectively the initial particle proper
time and the causal retarded proper time (see Paper I), and $\mathbf{X}_{H}$
the Hamiltonian vector field%
\begin{equation}
\mathbf{X}_{H}\left( \mathbf{y},\left[ r\right] \right) \equiv \left\{ \frac{%
\partial H_{eff}\left( r,P,\left[ r\right] \right) }{\partial P_{\mu }},-%
\frac{\partial H_{eff}\left( r,P,\left[ r\right] \right) }{\partial r^{\mu }}%
\right\} .  \label{hamvect}
\end{equation}%
Denoting by $\mathbf{y}\left( s\right) =\mathbf{\chi }\left( \mathbf{y}%
_{0},\left\{ \mathbf{y}_{s_{0}^{\prime }},\forall s_{0}^{\prime }\in
I_{s_{0},s_{0}-s_{ret}}\right\} ,s-s_{0}\right) $ the formal solution of the
problem (\ref{initial pro}), in the reminder we shall assume that the map%
\begin{equation}
\mathbf{y}_{0}\rightarrow \mathbf{y}\left( s\right)  \label{map}
\end{equation}%
is a diffeomeorphism of class $C^{k}$, with $k\geq 1$.

Based on these results, the Hamiltonian formulation is provided by the
following theorem.

\bigskip

\textbf{THM.4 - Non-local variational and effective Hamiltonian functions
for the non-rotating particle}

\emph{Given validity of THMs. 1-3, it follows that:}

\emph{T4}$_{1}$\emph{) The RR equation (\ref{RR}) for a non-rotating and
spherically-symmetric charged particle admits the non-local Hamiltonian
system }$\left\{ \mathbf{y}\equiv (r^{\mu }\mathbf{,}p_{\mu })\mathbf{,}%
H_{1}\right\} $\emph{. Here }$p_{\mu }$ \emph{and }$H_{1}\equiv H_{1}\left(
r,p,\left[ r\right] \right) $\emph{\ are respectively the canonical momentum
(\ref{canmom}) defined with respect to the variational Lagrangian }$L_{1}^{%
\text{ }}$ \emph{given in Eq.(\ref{EXTREMAL LAGRANGIAN}), and the
corresponding non-local variational Hamiltonian (\ref{canh}) defined as the
Legendre transformation of }$L_{1}$\emph{. In particular, the variational
non-local Hamiltonian (\ref{canh}) is identified with}%
\begin{equation}
H_{1}\left( r,p,\left[ r\right] \right) \equiv \frac{1}{2m_{o}c}\left(
p_{\mu }-\frac{q}{c}A_{\mu }\right) \left( p^{\mu }-\frac{q}{c}A^{\mu
}\right) ,  \label{NON-LOCAL HAM}
\end{equation}%
\emph{where }$A_{\mu }$\emph{\ is the total EM 4-potential}%
\begin{equation}
A_{\mu }\left( r,\left[ r\right] \right) \equiv \overline{A}_{\mu
}^{(ext)}\left( r\right) +\overline{A}_{\mu }^{(self)}\left( r,\left[ r%
\right] \right) ,  \label{atot1}
\end{equation}%
\emph{and from Eq.(\ref{LAGRANGIAN-SELF-EM}) }$\overline{A}_{\mu }^{(self)}$%
\emph{\ is the functional}%
\begin{equation}
\overline{A}_{\mu }^{(self)}\left( r,\left[ r\right] \right) \equiv
2q\int_{1}^{2}dr_{\mu }^{\prime }\delta (\widetilde{R}^{\mu }\widetilde{R}%
_{\mu }-\sigma ^{2}).  \label{atot2}
\end{equation}

\emph{T4}$_{2}$\emph{) There exist} $P_{\mu }$ \emph{and} $H_{eff},$ \emph{%
defined respectively by Eqs.(\ref{pleff}) and (\ref{heff}),\textit{\ }such
that}%
\begin{equation}
H_{eff}\left( r,P,\left[ r\right] \right) \equiv \frac{1}{2m_{o}c}\left(
P_{\mu }-\frac{q}{c}A_{\left( eff\right) \mu }\right) \left( P^{\mu }-\frac{q%
}{c}A_{\left( eff\right) }^{\mu }\right) ,  \label{EFFECGIVE HAMILTONIAN}
\end{equation}%
\emph{with }$A_{\left( eff\right) \mu }$\emph{\ the non-local effective EM
4-potential}%
\begin{equation}
A_{\left( eff\right) \mu }\left( r,P\right) \equiv \overline{A}_{\mu
}^{(ext)}\left( r\right) +2\overline{A}_{\mu }^{(self)}\left( r,\left[ r%
\right] \right)
\end{equation}%
\emph{and }$\overline{A}_{\mu }^{(self)}$ \emph{defined in Eq.(\ref{atot2}).}

\emph{T4}$_{3}$\emph{) The effective and variational Hamiltonian functions }$%
H_{eff}$\emph{\ and }$H_{1}$\emph{\ coincide when expressed in terms of the
4-velocity }$\frac{dr^{\mu }(s)}{ds}$\emph{.}

\emph{Proof}\ - The proof of T4$_{1}$ and T4$_{2}$ follows immediately by
applying THMs.1 and 2 with the variational Lagrangian $L_{1}$ given by Eq.(%
\ref{EXTREMAL LAGRANGIAN}). In particular, this yields%
\begin{equation}
p_{\mu }=m_{o}c\frac{dr_{\mu }(s)}{ds}+\frac{q}{c}\left[ \overline{A}_{\mu
}^{(ext)}+\overline{A}_{\mu }^{(self)}\right]  \label{p}
\end{equation}%
and%
\begin{equation}
P_{\mu }=m_{o}c\frac{dr_{\mu }(s)}{ds}+\frac{q}{c}\left[ \overline{A}_{\mu
}^{(ext)}+2\overline{A}_{\mu }^{(self)}\right] .  \label{pp}
\end{equation}%
The corresponding Legendre transformations then provide respectively Eq.(\ref%
{NON-LOCAL HAM}) and Eq.(\ref{EFFECGIVE HAMILTONIAN}). Finally, by direct
substitution of Eq.(\ref{p}) into Eq.(\ref{NON-LOCAL HAM}) and Eq.(\ref{pp})
into Eq.(\ref{EFFECGIVE HAMILTONIAN}), one obtains that%
\begin{equation}
H_{eff}=H_{1}=\frac{m_{o}c}{2}\frac{dr_{\mu }(s)}{ds}\frac{dr^{\mu }(s)}{ds},
\end{equation}%
which proves also the last statement.

\textbf{Q.E.D.}

\bigskip

We remark that the Hamilton equation in standard form expressed in terms of $%
H_{eff}$ and $P_{\mu }$ are differential equations of delay-type, as a
consequence of the non-local dependencies appearing in $H_{eff}$ which are
characteristic of the RR phenomenon. In this case, for the well-posedness of
the solution the initial conditions in the interval $I=\left[
s_{0}-s_{ret},s_{0}\right] $ must be defined, with $s_{0}$ the initial
proper time and $s_{ret}$ a suitable retarded time. However, if the
assumption of inertial motion in the proper time interval $I_{0}=\left[
-\infty ,s_{0}\right] $ holds, then the mapping%
\begin{equation}
T_{s_{0},s}:\mathbf{y}_{0}\equiv \mathbf{y}\left( s_{0}\right) \rightarrow
\mathbf{y}\left( s\right) \equiv T_{s_{0},s}\mathbf{y}_{0},  \label{dinsis}
\end{equation}%
with $\mathbf{y}=\left( r^{\mu },P_{\mu }\right) $, defines a classical
dynamical system (see Paper I), and this dynamical system is Hamiltonian.

\bigskip

\section{A Hamiltonian asymptotic approximation for the RR equation}

In this section we intend to carry out in detail a comparison of the present
approach for extended particles with the customary point-particle treatments
leading to the LAD and LL equations. For this purpose, asymptotic
approximations of the exact RR self-force (\ref{COVARIANT FORM}) are
investigated.

The issue has been partially discussed in Paper I. As pointed out therein,
an asymptotic approximation of the exact RR equation (\ref{RR}) can be
obtained in validity of the \emph{short delay-time ordering}, namely
requiring%
\begin{equation}
0<\epsilon \equiv \left\vert \frac{s_{ret}}{s}\right\vert \ll 1,
\label{ordering -1}
\end{equation}%
where $s_{ret}=s-s^{\prime }$, with $s$ and $s^{\prime }$ denoting
respectively the present and retarded particle proper times. This permits
two different possible strategies, respectively based on Taylor expansions
performed with respect to $s$ (\textit{present-time expansion}) or $%
s^{\prime }$ (\textit{retarded-time expansion}). In particular, adopting the
present-time expansion for the RR self-force (\ref{COVARIANT FORM}), the
delay-type ODE (\ref{RR}) can be reduced, in principle, to an infinite-order
differential equation. Instead, by truncating the Taylor expansion to
first-order in $\epsilon $, ignoring mass-renormalization terms and taking
the point-particle limit $\sigma \rightarrow 0$, in this way the customary
expression for the LAD equation is recovered (see THM.3 of Paper I).

As remarked in the Introduction, the resulting asymptotic approximation
(given by the LAD equation) is non-variational and therefore
non-Hamiltonian. In addition, contrary to the exact RR equation obtained
here, the LAD equation, as well as the related LL approximation, both fail
in the transient time intervals occurring when the external EM field acting
on the particle is turned on and off. To elucidate this point, let us
consider the dynamics of a charged particle which is in inertial motion in
the past for all $s<s_{0}$ and from $s=s_{0}$ is subject to the action of an
external EM field. Then, by construction, it is immediate to show that in
the transient time interval $I_{0}=\left[ s_{0},s_{0}+s_{ret}\right] $ the
exact RR self-force (\ref{RR}) is manifestly identically zero. In fact, in
the case of inertial motion in the past (namely $u_{\mu }(s^{\prime
})=const. $) the RR self-force vanishes in such a time interval (see THM.1
in Paper I). In contrast, both the LAD and LL equations predict incorrectly
a non-vanishing RR self-force. The same kind of inconsistency (for the LAD
and LL equations) arises when the analogous transient time interval
corresponding to the turning-off of the external EM field is considered \cite%
{Dorigo2008a}.

Therefore, the issue arises whether an alternative asymptotic approximation
can be determined (for the exact RR equation) which simultaneously:

1) overcomes this deficiency, by taking into account consistently
relativistic finite delay-time effects characteristic of the RR phenomenon;

2) is variational and admits a standard Hamiltonian formulation.

In this section we propose a solution to this problem, by performing a
retarded-time expansion, which provides an alternative to the LAD and LL
equations.

\bigskip

\subsection{The Hamiltonian approximation}

For definiteness, let us assume that the external force acting on the
particle is non-vanishing only in a finite proper-time interval $I\equiv %
\left[ s_{0},s_{1}\right] $. Then, in validity of the ordering (\ref%
{ordering -1}), we require that the external EM force is slowly varying in
the sense that, denoting $r^{\prime }\equiv r^{\mu }\left( s^{\prime
}\right) $ and $r\equiv r^{\mu }\left( s\right) $,%
\begin{eqnarray}
\overline{F}_{\mu \nu }^{(ext)}\left( r^{\prime }\right) -\overline{F}_{\mu
\nu }^{(ext)}\left( r\right) &\sim &O\left( \epsilon \right) ,
\label{smooth1} \\
\left( \overline{F}_{\mu \nu }^{(ext)}\left( r^{\prime }\right) -\overline{F}%
_{\mu \nu }^{(ext)}\left( r\right) \right) _{,h} &\sim &O\left( \epsilon
\right) , \\
\left( \overline{F}_{\mu \nu }^{(ext)}\left( r^{\prime }\right) -\overline{F}%
_{\mu \nu }^{(ext)}\left( r\right) \right) _{,hk} &\sim &O\left( \epsilon
\right) .  \label{smooth3}
\end{eqnarray}

Then, the retarded-time Hamiltonian approximation of the RR equation is
obtained by performing a Taylor expansion in a neighborhood of $s^{\prime }$%
. The result is summarized by the following theorem.

\bigskip

\textbf{THM.5 - First-order, short delay-time Hamiltonian approximation
(retarded-time expansion).}

\emph{Given validity of the asymptotic ordering (\ref{ordering -1}) and the
smoothness assumptions (\ref{smooth1})-(\ref{smooth3}) for the external EM
field, neglecting corrections of order }$\epsilon ^{n},$ \emph{with} $n\geq
1 $ \emph{(first-order approximation), the following results hold:}

\emph{T5}$_{1}$\emph{) The vector field}%
\begin{equation}
G_{\mu }\equiv \frac{q}{c}\overline{F}_{\mu k}^{\left( self\right) }\left(
r\left( s\right) ,r\left( s^{\prime }\right) \right) \frac{dr^{k}(s)}{ds}
\end{equation}%
\emph{appearing in Eq.(\ref{RR}) can be approximated in a neighborhood of }$%
s^{\prime }$ \emph{as}%
\begin{equation}
g_{\mu }\left( r\left( s^{\prime }\right) \right) =\left\{ -m_{oEM}c\frac{d}{%
ds^{\prime }}u_{\mu }\left( s^{\prime }\right) +g_{\mu }^{\prime }\left(
r\left( s^{\prime }\right) \right) \right\} ,  \label{asymp}
\end{equation}%
\emph{\ to be referred to as retarded-time Hamiltonian approximation, in
which the first term on the rhs identifies a retarded mass-correction term, }%
$m_{oEM}\equiv \frac{q^{2}}{c^{2}\sigma }$ \emph{denoting the leading-order
EM mass. Finally, }$g_{\mu }^{\prime }$\emph{\ is the 4-vector}%
\begin{equation}
g_{\mu }^{\prime }\left( r\left( s^{\prime }\right) \right) =-\frac{1}{3}%
\frac{q^{2}}{c}\left[ \frac{d^{2}}{ds^{\prime 2}}u_{\mu }\left( s^{\prime
}\right) -u_{\mu }(s^{\prime })u^{k}(s^{\prime })\frac{d^{2}}{ds^{\prime 2}}%
u_{k}\left( s^{\prime }\right) \right] .
\end{equation}

\emph{T5}$_{2}$\emph{) The corresponding RR equation, obtained replacing }$%
G_{\mu }$ \emph{with the asymptotic approximation }$g_{\mu }$ \emph{(\ref%
{asymp}), is variational, Lagrangian and admits a standard Lagrangian form.
Let us denote with }$r_{0}^{\prime }\equiv r_{0}\left( s^{\prime }\right) $
\emph{the extremal particle world-line at the retarded proper time }$%
s^{\prime }$\emph{. Then, in this approximation the corresponding asymptotic
variational Lagrangian and effective Lagrangian functions coincide. Both are
defined in terms of the asymptotic approximation }$%
L_{C,asym}^{(self)}(r,r_{0}^{\prime }),$\emph{\ replacing }$L_{C}^{(self)}$%
\emph{. To leading-order in }$\epsilon $\emph{, this is found to be}%
\begin{equation}
L_{C,asym}^{(self)}(r,r_{0}^{\prime })=g_{\mu }\left( r_{0}^{\prime }\right)
r^{\mu }.
\end{equation}%
\emph{T5}$_{3}$\emph{) The asymptotic approximation given by Eq.(\ref{asymp}%
) is also Hamiltonian. The asymptotic variational and effective Hamiltonian
functions coincide and are given by}%
\begin{equation}
H_{1,asym}=p_{\mu }\frac{dr^{\mu }}{ds}-L_{1,asym}
\end{equation}%
\emph{with}%
\begin{equation}
L_{1,asym}(r,r_{0}^{\prime
})=L_{M}(r)+L_{C}^{(ext)}(r)+L_{C,asym}^{(self)}(r,r_{0}^{\prime }),
\end{equation}%
\emph{and now}%
\begin{equation}
p_{\mu }=\frac{\partial L_{1,asym}}{\partial \frac{dr_{\mu }(s)}{ds}}.
\end{equation}

\emph{Proof} - T5$_{1}$) The proof can be carried out starting from Eq.(\ref%
{RR}) and performing explicitly the Taylor expansion in a neighborhood of $%
s^{\prime }\equiv s-s_{ret}$. For a generic analytic function $f\left(
s\right) $, this yields the power series of the form%
\begin{equation}
f(s)=\sum\limits_{k=0}^{\infty }\frac{(s-s^{\prime })^{k}}{k!}\frac{%
d^{k}f(s^{\prime })}{ds^{k}}.  \label{Taylor-series2}
\end{equation}%
In particular, for the 4-vectors $\frac{dr_{\mu }(s)}{ds}$ and $\widetilde{R}%
^{k}$ one obtains respectively the asymptotic approximations%
\begin{equation}
\frac{dr_{\mu }(s)}{ds}\cong \frac{dr_{\mu }\left( s^{\prime }\right) }{%
ds^{\prime }}+(s-s^{\prime })\frac{d^{2}r_{\mu }\left( s^{\prime }\right) }{%
ds^{\prime 2}}+\frac{(s-s^{\prime })^{2}}{2}\frac{d^{3}r_{\mu }\left(
s^{\prime }\right) }{ds^{\prime 3}}+O\left( \epsilon ^{3}\right)
\end{equation}%
and%
\begin{equation}
\widetilde{R}^{k}\cong (s-s^{\prime })\frac{dr^{k}\left( s^{\prime }\right)
}{ds^{\prime }}+\frac{(s-s^{\prime })^{2}}{2}\frac{d}{ds^{\prime }}%
u^{k}\left( s^{\prime }\right) +\frac{(s-s^{\prime })^{3}}{6}\frac{d^{2}}{%
ds^{\prime 2}}u^{k}\left( s^{\prime }\right) +O\left( \epsilon ^{4}\right) ,
\end{equation}%
while for the time delay $s-s^{\prime }\equiv s_{ret}$ the leading-order
expression%
\begin{equation}
s-s^{\prime }\cong \sigma +O\left( \epsilon ^{2}\right)
\end{equation}%
holds. By substituting these expansions in Eq.(\ref{COVARIANT FORM}), the
asymptotic solution given by Eq.(\ref{asymp}) can be recovered.

T5$_{2}$)-T5$_{3}$) The proof follows by first noting that $%
L_{C,asym}^{(self)}$ contributes to the Euler-Lagrange equations only in
terms of the local dependence in terms of $r$. Then, in this approximation
the canonical momentum becomes%
\begin{equation}
p_{\mu }=m_{0}c\frac{dr_{\mu }(s)}{ds}+\frac{q}{c}\overline{A}_{\mu
}^{(ext)}(r)=P_{\mu },
\end{equation}%
while the asymptotic Hamiltonian reduces to%
\begin{equation}
H_{1,asym}\left( r,p,r_{0}^{\prime }\right) =\frac{1}{2m_{o}c}\left( p_{\mu
}-\frac{q}{c}\overline{A}_{\mu }^{(ext)}(r)\right) \left( p^{\mu }-\frac{q}{c%
}\overline{A}_{\mu }^{(ext)\mu }(r)\right) +g_{\mu }\left( r_{0}^{\prime
}\right) r^{\mu }.
\end{equation}%
The corresponding Lagrangian and Hamiltonian equations manifestly coincide
with Eq.(\ref{RR}) once the approximation (\ref{asymp}) is invoked for the
vector field $G_{\mu }$.

\textbf{Q.E.D.}

\bigskip

\subsection{Discussion and comparisons with point-particle treatments}

The asymptotic Hamiltonian approximation, here pointed out for the first
time (see THM.5), preserves the basic physical properties of the exact RR
force (\ref{RR}). In fact in both cases, the RR force:

1) is non-local, depending on the past history of the finite-size charged
particle;

2) admits a variational formulation;

3) is both Lagrangian and Hamiltonian;

4) satisfies the Einstein Causality Principle and, when applicable, the
Newton Principle of Determinacy (see also Paper I);

5) describes correctly the transient time intervals in which the external
force is turned on and off (sudden force).

\bigskip

For these reasons, physical comparisons based on the retarded-time
Hamiltonian asymptotic approximation are meaningful. In particular, here we
remark that the present approach departs in several ways with respect to
point-particle treatments based on the LAD and LL equations. More precisely:

1) The same type of asymptotic ordering is imposed, which is based on the
short delay-time ordering (\ref{ordering -1}). However, in contrast with the
LAD and LL equations, the expansion adopted in THM.5 and leading to the
retarded-time Hamiltonian approximation can \emph{only} be performed based
on the knowledge of the exact RR force for finite-size particles.

2) Unlike the LAD and LL equations, the asymptotic Hamiltonian approximation
carries the information of the past dynamical history of the charged
particle through the retarded time $s^{\prime }$. Therefore, the dynamical
equation written adopting the approximation (\ref{asymp}) is still a
delay-type second-order ODE. The construction of its general solution
becomes trivial in this case, since the self-force is considered as an
explicit source term evaluated at proper time $s^{\prime }$.

3) The asymptotic approximation provided by Eq.(\ref{asymp}) cannot be
regarded as a point-particle limit. In fact, the retarded mass-correction
term would diverge in this limit.

4) The exact RR equation satisfies identically by construction the kinematic
constraint $u_{\mu }u^{\mu }=1$. The same constraint is satisfied to
leading-order in $\epsilon $ also both by the retarded and present-time
asymptotic expansions (and hence also the LAD equation).

5) The variational principle introduced in THM.5 is subject to the
constraint that the past history is considered prescribed in terms of the
extremal world-line. This requirement is consistent with the initial
conditions for the RR equation, which is a delay-type ODE depending only on
the past history of the particle. This requires that the world-line
trajectory is prescribed in the past, namely in the time interval $I=\left[
-\infty ,s_{0}\right] $. Since, however, the initial proper time $s_{0}$ is
arbitrary, it follows that $r\left( s\right) $ can be considered prescribed
also in the time interval $I^{\prime }=\left[ -\infty ,s^{\prime }\right] $.
In particular, if for all $s<s_{0}$ the motion is assumed to be inertial,
the initial-value problem associated to the RR equation written in terms of
the retarded asymptotic self-force (\ref{asymp}) is well-posed, in the sense
of the standard Newton Principle of Determinism, as discussed in Paper I
(see in particular THM.4 presented there and dealing with the existence and
uniqueness of solutions for the exact RR equation).

6) One might think that the same type of constrained variational principle,
of the kind adopted in THM.5, could be adopted also for the exact RR
equation. However, this belief is wrong. In fact, since the variational
functional (\ref{EXTREMAL LAGRANGIAN}) is symmetric with respect to the
local and non-local world-line trajectories, there is no distinction between
past and future. Since future cannot be prescribed, such a constrained
variational principle for the exact equation is forbidden. On the contrary,
the extremal RR equation (\ref{RR}) is obtained by imposing also the
Einstein Causality Principle, and therefore it depends only on the past
history.

7) Despite some formal similarities between the retarded-time Hamiltonian
approximation versus the corresponding LAD and LL equations, the latter
cannot be recovered even in the framework of some kind of constrained
variational principle. In fact this would require to consider prescribed for
example, second or higher-order proper-time derivatives of the particle
position vector (namely the acceleration and its derivatives). This
viewpoint is manifestly unacceptable, because it would amount to constraint
the present state of the particle at proper time $s$.

8) The previous argument justifies, in turn, the introduction of the short
delay-time asymptotic approximation given in THM.5. This is performed
directly on the RR force, namely the 4-vector $G_{\mu }$ entering the RR
equation itself. In this way the variational character of the RR problem is
preserved. It follows that the corresponding variational functional as well
as the Lagrangian and Hamiltonian functions for the asymptotic RR equation
are constructed only \textquotedblleft a posteriori\textquotedblright .

9) Another advantage of the new representation (\ref{asymp}) with respect to
the customary LAD and LL equations is that it permits the approximate
treatment of the solution also in the transient\ time intervals after the
turning-on or the turning-off of the external EM field. In particular, in
contrast to the LAD and LL equations, it predicts a vanishing RR self-force
in the turning-on transient phase $I_{0}=\left[ s_{0},s_{0}+s_{ret}\right] $.

10) Finally, it should be remarked that the retarded asymptotic self-force (%
\ref{asymp}) \emph{cannot} be trivially obtained from the corresponding
local asymptotic representation performed at proper time $s$ and leading to
the LAD equation by simply exchanging $s$ with $s^{\prime }$ (or by a
further Taylor expansion). Indeed, the relationship between the two can only
be established based on the exact form of the self-force.

\bigskip

\section{Collisionless relativistic kinetic theory for the EM RR effect -
Canonical formalism}

In this section we proceed with the construction of the relativistic
classical statistical mechanics (CSM) for a collisionless plasma with the
inclusion of the EM RR effect. In particular we shall prove that the
mathematical formalism introduced in the previous sections to deal with
symmetric non-local interactions allows one to obtain a convenient
formulation for the kinetic theory describing such a system and for the
corresponding fluid representation. The derivation is based on the property
of a symmetric non-local system represented by a finite-size charged
particle of being Hamiltonian with respect to $P_{\mu }$ and $H_{eff}$.

In view of the peculiar features of the non-local RR phenomenon and the
related delay-type differential Hamiltonian equations, it is instructive
here to adopt an axiomatic formulation of the CSM for relativistic systems
with the inclusion of such an effect. We shall assume that the latter are
represented by a system of classical finite-size charged particles subject
only to the action of a mean-field external EM force and a non-local
self-interaction. We intend to show that, using the Hamiltonian
representation in standard form given above, the explicit form of the
relativistic Vlasov kinetic equation can be obtained for the kinetic
distribution function describing the statistical dynamics of such a system.
Therefore, the problem is reduced to a Vlasov-Maxwell description for a
continuous distribution of relativistic charged particles.

For definiteness, let us consider the non-local Hamiltonian dynamical system
in standard form $\left\{ \mathbf{y},H_{eff}\right\} $ given above. This is
characterized by the superabundant state vector $\mathbf{y}=\left( r^{\mu
},P_{\mu }\right) $ spanning the extended 8th-dimensional phase-space $%
\Gamma $ and with essential state variables $\mathbf{y}_{1}\left( \mathbf{y}%
\right) $ spanning the 6th-dimensional reduced phase-space $\Gamma _{1}$.
Introducing the \textit{global proper time} $\widehat{s}$, $\Gamma
_{1}\left( \widehat{s}\right) $ is defined as%
\begin{equation}
\Gamma _{1}\left( \widehat{s}\right) \equiv \left\{ \mathbf{y}:\mathbf{y}\in
\Gamma ,\ \left\vert u\right\vert =1,\ s\left( y\right) =\widehat{s},\
ds\left( y\right) =\sqrt{g_{\mu \nu }dr^{\mu }dr^{\nu }}\right\} ,
\end{equation}%
where $\left\vert u\right\vert \equiv \sqrt{u^{\alpha }u_{\alpha }}$ and $%
s\left( y\right) $ is the world-line proper time uniquely associated to any $%
\mathbf{y}$. By assumption, $\Gamma _{1}\left( \widehat{s}\right) $ is an
invariant set, i.e., $\Gamma _{1}\left( \widehat{s}\right) =\Gamma _{1}$ for
any $\widehat{s}\in
\mathbb{R}
$. Next, let us consider the Hamiltonian flow $T_{s_{0},s}$ defined in Eq.(%
\ref{dinsis}). By construction the dynamical system is autonomous, namely
the flow is of the form%
\begin{equation}
T_{s_{0},s}\mathbf{y}_{0}\equiv \chi \left( \mathbf{y}_{0},s-s_{0}\right) .
\end{equation}%
The existence of the dynamical system $T_{s_{0},s}$ for the state $\mathbf{y}%
\left( s\right) $ has been proved in Paper I. This requires that in the
proper time interval $I_{0}=\left[ -\infty ,s_{0}\right] $ the motion of
each charged particle is inertial, namely the external EM field vanishes in
the same interval. As a result of Eq.(\ref{dinsis}), any point in the
phase-space $\Gamma $ spanned by $\mathbf{y}$ or $\mathbf{y}_{0}$ is
associated to a unique phase-space trajectory, namely such that $\mathbf{y}=%
\mathbf{y}\left( s\right) $, for any $\mathbf{y}\in \Gamma $. Due to (\ref%
{dinsis}) there exists necessarily $\mathbf{y}_{0}\equiv \mathbf{y}\left(
s_{0}\right) $ which is mapped in $\mathbf{y}\left( s\right) .$ Viceversa,
for any $s\in
\mathbb{R}
$ there exists a unique $\mathbf{y}=\mathbf{y}\left( s\right) $. However, we
notice here that for the axiomatic formulation of the CSM for the RR problem
the assumption of existence of the dynamical system $T_{s_{0},s}$ is not a
necessary condition. In fact, it is immediate to prove that the minimal
requirement is actually provided only by the existence of the
diffeomeorphism (\ref{map}) defined above.

Now, for a prescribed $\widehat{s}_{0}\in
\mathbb{R}
$ let us consider the set $B\left( \widehat{s}_{0}\right) \subseteq \Gamma
_{1}$, with $B\left( \widehat{s}_{0}\right) $ an ensemble of states $\mathbf{%
y}_{0},$ each one prescribed at the initial proper time $s_{0}=\widehat{s}%
_{0}$. Its image generated at any $s=\widehat{s}\in
\mathbb{R}
$ by the flow $T_{s_{0},s}$, for each trajectory, is%
\begin{equation}
B\equiv B\left( s\right) \equiv T_{s_{0},s}B\left( s_{0}\right) ,
\end{equation}%
where $s$ and $s_{0}$ denote now the \textit{global proper times} $\widehat{s%
}$ and $\widehat{s}_{0}$.

We introduce the following axioms.

\emph{AXIOM \#1:\ Probability on }$K\left( \Gamma \right) $.

Let $K\left( \Gamma _{1}\right) $ be a family of subsets of $\Gamma _{1}$
which are $L$-measurable. We define the probability of $B\left( s\right) \in
K\left( \Gamma _{1}\right) $ as the function%
\begin{equation}
P\left( B\right) :K\left( \Gamma _{1}\right) \rightarrow \left[ 0,1\right]
\end{equation}%
such that it satisfies the constraints%
\begin{eqnarray}
P\left( \Gamma _{1}\right) &=&1, \\
P\left( \varnothing \right) &=&0, \\
P\left( \cup _{i\in N}B_{i}\right) &=&\sum_{i=0}^{\infty }P\left(
B_{i}\right) ,
\end{eqnarray}%
with $\left\{ B_{i}\in K\left( \Gamma _{1}\right) ,i\in N\right\} $ being an
arbitrary family of separate sets of $K\left( \Gamma _{1}\right) $.

\emph{AXIOM \#2:\ Probability density.}

For any $B\left( s\right) \in K\left( \Gamma _{1}\right) $ and for any state
$\mathbf{y\equiv }\left( r^{\mu },P_{\mu }\right) $ there exists a unique
probability density $\rho \left( \mathbf{y}\right) >0$ on $\Gamma _{1}$ such
that%
\begin{equation}
P\left( B\left( s\right) \right) =\int_{\Gamma }d\mathbf{y}\rho \left(
\mathbf{y}\right) \delta \left( \left\vert u\right\vert -1\right) \delta
\left( s-s\left( y\right) \right) \delta _{B\left( s\right) }\left( \mathbf{y%
}\right) ,  \label{intphasesp}
\end{equation}%
where $d\mathbf{y}=dr^{\mu }dP_{\mu }$ is the canonical measure on $\Gamma $
and $\delta _{B\left( s\right) }\left( \mathbf{y}\right) $ is the
characteristic function of $B\left( s\right) $. Furthermore, $s\left(
y\right) $ is a particle world-line proper time, while $s\equiv s_{0}+\Delta
s$, with $\Delta s$ an \textit{invariant proper time interval} independent
of $s_{0}$. We notice that $s\left( y\right) $ can be equivalently
parametrized in terms of the observer's coordinate time $r^{0}$, namely:%
\begin{equation}
ds\left( y\right) \equiv dr^{0}\sqrt{g_{\mu \nu }\frac{dr^{\mu }}{dr^{0}}%
\frac{dr^{\nu }}{dr^{0}}}.
\end{equation}

\emph{AXIOM \#3:\ Equiprobability.}

Then, the equiprobability condition requires that, for all $B\left(
s_{0}\right) $ and for all $s,s_{0}\in I\subseteq
\mathbb{R}
$,%
\begin{equation}
P\left( B\left( s\right) \right) =P\left( B\left( s_{0}\right) \right) .
\end{equation}

\bigskip

We remark that in the integral (\ref{intphasesp}) the two Dirac-delta
functions can be interpreted as physical realizability conditions, required
to reduce the dimension of the volume element $d\mathbf{y}$ defined on the
extended phase-space $\Gamma $.

We can now introduce the following theorem, concerning the validity of the
Liouville equation for $\rho \left( \mathbf{y}\right) $.

\bigskip

\textbf{THM.6 - Relativistic Liouville equation for }$\rho \left( \mathbf{y}%
\right) $.

\emph{Given a Hamiltonian system }$\left\{ \mathbf{y},H_{eff}\right\} $
\emph{and imposing the validity of Axioms \#1-\#3, it follows that the
probability density }$\rho \left( \mathbf{y}\left( s\right) \right) $ \emph{%
is a constant of motion, namely for any} $s,s_{0}\in
\mathbb{R}
$ \emph{(to be intended now as world-line proper times) and for any }$%
\mathbf{y}_{0}\in \Gamma $%
\begin{equation}
\rho \left( \mathbf{y}\left( s\right) \right) =\rho \left( \mathbf{y}%
_{0}\right) ,
\end{equation}%
\emph{to be referred to as the integral Liouville equation. This can also be
written equivalently as}%
\begin{equation}
\frac{d}{ds}\rho \left( \mathbf{y}\left( s\right) \right) =0,  \label{liouv}
\end{equation}%
\emph{to be referred to as the differential Liouville equation. As a
consequence, introducing the kinetic distribution function (KDF) }$f\left(
\mathbf{y}\right) $%
\begin{equation}
f\left( \mathbf{y}\right) \equiv \rho \left( \mathbf{y}\right) N,
\end{equation}%
\emph{with }$N$\emph{\ being the total number of particles in the
configuration space of }$B\subseteq K\left( \Gamma \right) $\emph{, it
follows that also }$f\left( \mathbf{y}\right) $\emph{\ satisfies the
Liouville equation (\ref{liouv}).}

\emph{Proof}\ - We first notice that, from Axiom \#1, by changing the
integration variables we can write Eq.(\ref{intphasesp}) as%
\begin{eqnarray}
P\left( B\left( s\right) \right) &=&\int_{\Gamma }d\mathbf{y}\rho \left(
\mathbf{y}\right) \delta \left( \left\vert u\right\vert -1\right) \delta
\left( s-s\left( y\right) \right) \delta _{B\left( s\right) }\left( \mathbf{y%
}\right) =  \notag \\
&=&\int_{\Gamma }d\mathbf{y}_{0}\left\vert \frac{\partial \mathbf{y}\left(
s\right) }{\partial \mathbf{y}_{0}}\right\vert \rho \left( \mathbf{y}\left(
s\right) \right) \delta \left( \left\vert u\right\vert -1\right) \delta
\left( s-s\left( y\right) \right) \delta _{B\left( s_{0}\right) }\left(
\mathbf{y}\left( s_{0}\right) \right) ,
\end{eqnarray}%
with $\left\vert \frac{\partial \mathbf{y}\left( s\right) }{\partial \mathbf{%
y}_{0}}\right\vert $ being the Jacobian of the variable transformation from $%
\mathbf{y}\left( s\right) $ to $\mathbf{y}_{0}$. On the other hand, since
the system $\left\{ \mathbf{y},H_{eff}\right\} $ is Hamiltonian, it follows
identically that $\left\vert \frac{\partial \mathbf{y}\left( s\right) }{%
\partial \mathbf{y}_{0}}\right\vert =1$. Hence, invoking Axiom \#2 we can
write%
\begin{equation}
\int_{\Gamma }d\mathbf{y}_{0}\left[ \rho \left( \mathbf{y}\left( s\right)
\right) \delta \left( \left\vert u\right\vert -1\right) \delta \left(
s-s\left( y\right) \right) -\rho \left( \mathbf{y}_{0}\right) \delta \left(
\left\vert u_{0}\right\vert -1\right) \delta \left( s_{0}-s\left(
y_{0}\right) \right) \right] \delta _{B\left( s_{0}\right) }\left( \mathbf{y}%
\left( s_{0}\right) \right) =0,
\end{equation}%
from which it must be that%
\begin{equation}
\rho \left( \mathbf{y}\left( s\right) \right) \delta \left( \left\vert
u\right\vert -1\right) \delta \left( s-s\left( y\right) \right) =\rho \left(
\mathbf{y}_{0}\right) \delta \left( \left\vert u_{0}\right\vert -1\right)
\delta \left( s_{0}-s\left( y_{0}\right) \right) .  \label{Lio1}
\end{equation}%
On the other hand, by construction it follows that%
\begin{eqnarray}
\delta \left( \left\vert u\right\vert -1\right) &=&\frac{1}{\left\vert \frac{%
d\left\vert u\right\vert }{d\left\vert u_{0}\right\vert }\right\vert }\delta
\left( \left\vert u_{0}\right\vert -1\right) =\delta \left( \left\vert
u_{0}\right\vert -1\right) , \\
\delta \left( s-s\left( y\right) \right) &=&\frac{1}{\left\vert \frac{ds}{%
ds_{0}}\right\vert }\delta \left( s_{0}-s\left( y_{0}\right) \right) =\delta
\left( s_{0}-s\left( y_{0}\right) \right) .
\end{eqnarray}%
In fact, by definition the 4-velocity is normalized to 1 at all proper
times, so that $\left\vert \frac{d\left\vert u\right\vert }{d\left\vert
u_{0}\right\vert }\right\vert =1$. Furthermore, $s\equiv s_{0}+\Delta s$,
with $\Delta s$ being independent of the initial value $s_{0}$, and hence $%
\left\vert \frac{ds}{ds_{0}}\right\vert =1$ too.

Finally, because of these conclusions, from Eq.(\ref{Lio1}) it follows that%
\begin{equation}
\rho \left( \mathbf{y}\left( s\right) \right) =\rho \left( \mathbf{y}%
_{0}\right) ,
\end{equation}%
which represents the Liouville equation in integral form. By differentiating
with respect to $s$ the equivalent differential representation follows at
once. An analogous equation holds manifestly also for the KDF $f\left(
\mathbf{y}\right) $.

\textbf{Q.E.D.}

\bigskip

We conclude noting that, formally, the Liouville equation for non-local
Hamiltonian systems in standard form is analogous to that characterizing
local Hamiltonian systems. Such an equation can be viewed as a \textit{%
Vlasov equation} for a relativistic collisionless plasma, in which each
particle is subject only to the action of a mean-field EM interaction,
generated respectively by the external and the self\ EM Faraday tensors. By
definition, in this treatment the latter do not include retarded binary EM
interactions. It follows that, in terms of the Lagrangian equation (\ref%
{liouv}), the probability density $\rho \left( \mathbf{y}\left( s\right)
\right) $ is parametrized in terms of the single-particle phase-space
trajectory $\left\{ \mathbf{y}\left( s\right) ,s\in I\right\} $. Hence, it
advances in (proper) time $s$ by means of the canonical state $\mathbf{y}%
\left( s\right) $ as determined by the Hamiltonian equations of motion (\ref%
{initial pro}).

\bigskip

\subsection{Vlasov-Maxwell description}

To define a well-posed problem, the relativistic Vlasov equation (\ref{liouv}%
) must be coupled to the Maxwell equations, which determine the total EM
field produced by all the relevant sources. In particular, in order to
determine the external Faraday tensor $F_{\mu \nu }^{(ext)}$, the
corresponding EM 4-potential $A_{\nu }^{(ext)}$ must be determined. In the
Lorentz gauge, this is prescribed requiring it to be a solution of the
Maxwell equations%
\begin{equation}
\square A^{(ext)\mu }=\frac{4\pi }{c}j^{\left( ext\right) \mu }(r),
\label{m2}
\end{equation}%
where $j^{\left( ext\right) \mu }(r)$ is identified with the total current
density%
\begin{equation}
j^{\left( ext\right) \mu }(r)\equiv q\int d^{4}u\delta \left( \left\vert
u\right\vert -1\right) u^{\mu }f\left( \mathbf{y}\right) +j^{\left(
coils\right) \mu }(r).
\end{equation}%
Here, the first term is the Vlasov 4-current density, namely the velocity
moment of $f\left( \mathbf{y}\right) $ carrying the non-local phase-space
contributions which yield the collective field produced by the plasma. The
second term, instead, is produced by possible prescribed sources located
outside the plasma domain. Therefore, in the Vlasov-Maxwell description the
total EM 4-potential acting on a single particle must be considered as
represented by $A_{\nu }=A_{\nu }^{(ext)}+A_{\nu }^{(self)}$, where $A_{\nu
}^{(self)}$ is given by Eq.(\ref{intrepA}) and $A_{\nu }^{(ext)}$ is the
solution of Eq.(\ref{m2}).

Therefore, the dynamical evolution of the KDF along a single-particle
phase-space trajectory depends both explicitly, via $A_{\nu }^{(self)}$, and
implicitly, via the 4-current $j^{\left( ext\right) \mu }(r)$, on the whole
Faraday tensor $F_{\mu \nu }\equiv F_{\mu \nu }^{(ext)}+F_{\mu \nu
}^{(self)} $. In this way contributions which are non-local both in
configuration and phase-space are consistently included in the theory.

\bigskip

\section{Fluid moment equations}

We now proceed to compute explicitly the relativistic fluid moment equations
which follow from the Liouville equation. To this aim, the relativistic
Liouville equation is conveniently written as a PDE (Eulerian form)%
\begin{equation}
u^{\mu }\frac{\partial f\left( \mathbf{y}\right) }{\partial r^{\mu }}+G^{\mu
}\left( \mathbf{y}\right) \frac{\partial f\left( \mathbf{y}\right) }{%
\partial u_{\mu }}=0,  \label{euler-kinm}
\end{equation}%
where $G^{\mu }\left( \mathbf{y}\right) $ is defined by Eq.(\ref{RR}), or as
an ODE (Lagrangian form):%
\begin{equation}
\frac{dr^{\mu }}{ds}\frac{\partial f\left( \mathbf{y}\left( s\right) \right)
}{\partial r^{\mu }}+\frac{du_{\mu }}{ds}\frac{\partial f\left( \mathbf{y}%
\left( s\right) \right) }{\partial u_{\mu }}=0,  \label{lagr-kinm}
\end{equation}%
with $\mathbf{y}\left( s\right) $ being the phase-space trajectory of a
particle. Then, the relativistic fluid equations related to the Liouville
equation are defined as the following integrals over the momentum space:%
\begin{equation}
\int d^{4}u\delta \left( \left\vert u\right\vert -1\right) G\left[ u^{\mu }%
\frac{\partial f\left( \mathbf{y}\right) }{\partial r^{\mu }}+G^{\mu }\left(
\mathbf{y}\right) \frac{\partial f\left( \mathbf{y}\right) }{\partial u_{\mu
}}\right] =0.
\end{equation}%
Similarly, the corresponding fluid fields are defined as%
\begin{equation}
\int d^{4}u\delta \left( \left\vert u\right\vert -1\right) Gf\left( \mathbf{y%
}\right) ,
\end{equation}%
with $G=1,u^{\mu },u^{\mu }u^{\nu },...$ and $u^{\mu }$ is the 4-velocity.
In particular, we shall denote%
\begin{eqnarray}
n\left( r\right) &\equiv &\int d^{4}u\delta \left( \left\vert u\right\vert
-1\right) f\left( \mathbf{y}\right) ,  \label{den1} \\
N^{\mu }\left( r\right) &=&n\left( r\right) U^{\mu }\left( r\right) \equiv
\int d^{4}u\delta \left( \left\vert u\right\vert -1\right) u^{\mu }f\left(
\mathbf{y}\right) ,  \label{den2} \\
T^{\mu \nu }\left( r\right) &\equiv &\int d^{4}u\delta \left( \left\vert
u\right\vert -1\right) u^{\mu }u^{\nu }f\left( \mathbf{y}\right) ,
\label{den3}
\end{eqnarray}%
to be referred to as\textit{\ the number density, the 4-flow and the
stress-energy tensor.}

It is immediate to prove that the corresponding moment equations are as
follows.

\bigskip

\emph{Continuity equation}

For $G=1$ the Liouville equation provides the continuity equation%
\begin{equation}
\partial _{\mu }N^{\mu }\left( r\right) =0.  \label{continuity}
\end{equation}

\emph{Energy-momentum equation}

For $G=u^{\nu }$ the Liouville equation provides the energy-momentum equation%
\begin{equation}
\partial _{\mu }T^{\mu \nu }\left( r\right) =F_{\left( tot\right) }^{\nu \mu
}\left( r\right) N_{\mu }\left( r\right) ,  \label{momentum}
\end{equation}%
where, from Eq.(\ref{RR}) we have that%
\begin{equation}
F_{\left( tot\right) }^{\nu \mu }\left( r\right) \equiv \frac{q}{m_{o}c^{2}}%
\left[ \overline{F}^{\left( ext\right) \mu \nu }+\overline{F}^{\left(
self\right) \nu \mu }\right]
\end{equation}%
is the total EM force, with $\overline{F}^{\left( self\right) \nu \mu }$
containing the retarded non-local contributions arising from the EM RR
effect.

\bigskip

We remark the following properties.

1) As a consequence of the Hamiltonian formulation in standard form, the
fluid equations obtained from the kinetic equation with the inclusion of the
RR effect are formally the same as in the usual treatment for local systems.

2) The contribution of the RR effect to the fluid equations is contained
explicitly in the source term in the rhs of Eq.(\ref{momentum}), and also
implicitly in the definition of the fluid fields. In fact, by assumption,
the KDF is a function of the effective Hamiltonian state $\mathbf{y}\equiv
\left( r^{\mu },P_{\mu }\right) $, which depends on the retarded
self-potential. Hence, the fluid fields defined by Eqs.(\ref{den1})-(\ref%
{den3}) must be interpreted as the fluid fields of the plasma which is
emitting self-radiation and is therefore subject to the RR effect.

\bigskip

\subsection{The implicit contribution of the RR self-force}

It is worth discussing the features of the theory in connection with the
implicit contribution of the RR effect contained in the definition of the
fluid fields. In particular, here we show that such contribution can be made
explicit and an analytical asymptotic estimation of it can be give provide
some suitable assumptions are imposed on the physical system. This concerns
the case in which the contribution of the self-potential is small in
comparison with the external EM potential in the KDF. In these
circumstances, the exact KDF can be Taylor expanded as follows:%
\begin{equation}
f\left( \mathbf{y}\right) \simeq f\left( \mathbf{y}_{nc}\right) +\left(
\mathbf{y-y}_{nc}\right) \left. \frac{\partial f\left( \mathbf{y}\right) }{%
\partial \mathbf{y}}\right\vert _{\mathbf{y}=\mathbf{y}_{nc}}+...,
\label{series1}
\end{equation}%
where $\mathbf{y}_{nc}\equiv \left( r^{\mu },p_{\mu }\right) $ is the state
which is canonical in absence of the EM self-field. It is clear that, by
construction, only the canonical momenta are involved in this expansion,
since the configuration state is left unchanged by the presence of the
self-force. Therefore, from the form of the previous expansion it follows
that the first term of the series, namely $f\left( \mathbf{y}_{nc}\right) $,
does not contain any contribution from the RR self-field. Consider, for
simplicity, the Taylor series to first order. Then, the corresponding fluid
fields can be decomposed as follows:%
\begin{eqnarray}
n\left( r\right) &\simeq &n_{0}\left( r\right) +n_{1}\left( r\right) ,
\label{n1} \\
N^{\mu }\left( r\right) &\simeq &N_{0}^{\mu }\left( r\right) +N_{1}^{\mu
}\left( r\right) , \\
T^{\mu \nu }\left( r\right) &\simeq &T_{0}^{\mu \nu }\left( r\right)
+T_{1}^{\mu \nu }\left( r\right) ,  \label{t}
\end{eqnarray}%
where%
\begin{eqnarray}
n_{0}\left( r\right) &\equiv &\int d^{4}u\delta \left( \left\vert
u\right\vert -1\right) f\left( \mathbf{y}_{nc}\right) , \\
n_{1}\left( r\right) &\equiv &\int d^{4}u\delta \left( \left\vert
u\right\vert -1\right) \left( \mathbf{y-y}_{nc}\right) \left. \frac{\partial
f\left( \mathbf{y}\right) }{\partial \mathbf{y}}\right\vert _{\mathbf{y}=%
\mathbf{y}_{nc}}=\frac{2q}{c}\overline{A}_{\mu }^{(self)}\int d^{4}u\delta
\left( \left\vert u\right\vert -1\right) \left. \frac{\partial f\left(
\mathbf{y}\right) }{\partial P_{\mu }}\right\vert _{P_{\mu }=p_{\mu }},
\end{eqnarray}%
and similar definitions hold for the other two fluid fields.

To illustrate the procedure, let us consider, for example, the case of a
relativistic Maxwellian distribution of the form \cite{degroot}%
\begin{equation}
f_{M}\left( \mathbf{y}\right) \equiv \frac{1}{\left( 2\pi \hbar \right) ^{3}}%
\exp \left[ \frac{\mu -P^{\mu }U_{\mu }}{T}\right] ,  \label{max}
\end{equation}%
where $\mu $, $P^{\mu }$, $U_{\mu }$ and $T$ are respectively the chemical
potential, the canonical momentum and the fluid 4-velocity and temperature.
Then, in terms of the previous expansion, we obtain for the density%
\begin{eqnarray}
n_{0}\left( r\right) &\equiv &\frac{4\pi m^{2}cT}{\left( 2\pi \hbar \right)
^{3}}K_{2}\left( \frac{mc^{2}}{T}\right) \exp \left[ \frac{\mu }{T}-\frac{q}{%
c}\frac{\overline{A}_{\mu }^{(ext)}U^{\mu }}{T}\right] , \\
n_{1}\left( r\right) &\equiv &-\frac{2q}{c}\frac{\overline{A}_{\mu
}^{(self)}U^{\mu }}{T}n_{0}\left( r\right) ,
\end{eqnarray}%
with $K_{2}\left( \frac{mc^{2}}{T}\right) $ being the modified Bessel
function of the second kind. As can be seen, the effect of the RR self-field
appears only in $n_{1}\left( r\right) $ through the integral over the
non-local dependencies contained in the potential $\overline{A}_{\mu
}^{(self)}$. It follows that for a Maxwellian KDF the 4-flow $N^{\mu }\left(
r\right) $ can be written as%
\begin{equation}
N^{\mu }\left( r\right) \simeq \left[ n_{0}\left( r\right) +n_{1}\left(
r\right) \right] U^{\mu }\left( r\right) ,  \label{4flow_Max}
\end{equation}%
while the expansion terms of the stress-energy tensor $T^{\mu \nu }\left(
r\right) $ are given by%
\begin{eqnarray}
T_{0}^{\mu \nu }\left( r\right) &\equiv &\frac{1}{c^{2}}n_{0}eU^{\mu }U^{\nu
}-p_{0}\Delta ^{\mu \nu },  \label{t0} \\
T_{1}^{\mu \nu }\left( r\right) &\equiv &\frac{1}{c^{2}}n_{1}eU^{\mu }U^{\nu
}-p_{1}\Delta ^{\mu \nu }.  \label{t1}
\end{eqnarray}%
Here the notation is as in Ref.\cite{degroot}. Thus, $\Delta ^{\mu \nu }$ is
the projector operator $\Delta ^{\mu \nu }\equiv \eta ^{\mu \nu
}-c^{-2}U^{\mu }U^{\nu },$ $e$ is the energy per particle%
\begin{equation}
e=mc^{2}\frac{K_{3}\left( \frac{mc^{2}}{T}\right) }{K_{2}\left( \frac{mc^{2}%
}{T}\right) }-T
\end{equation}%
and from the definition of the pressure as $p=nT$ it follows that%
\begin{eqnarray}
p_{0}\left( r\right) &=&n_{0}\left( r\right) T, \\
p_{1}\left( r\right) &=&n_{1}\left( r\right) T=-\frac{2q}{c}\overline{A}%
_{\mu }^{(self)}U^{\mu }n_{0}\left( r\right) .
\end{eqnarray}

Finally, let us consider how the fluid equations are modified from the
introduction of the series expansion (\ref{series1}). Substituting the
relations (\ref{n1})-(\ref{t}) into the moment equations, for the continuity
equation we get%
\begin{equation}
\partial _{\mu }N_{0}^{\mu }\left( r\right) =-\partial _{\mu }N_{1}^{\mu
}\left( r\right) ,  \label{as1}
\end{equation}%
and for the momentum equation%
\begin{equation}
\partial _{\mu }T_{0}^{\mu \nu }=F_{\left( tot\right) }^{\nu \mu }N_{\mu
}-\partial _{\mu }T_{1}^{\mu \nu }.  \label{as2}
\end{equation}%
In this way, on the lhs we have isolated the terms of the \textquotedblleft
unperturbed fluid\textquotedblright , namely the physical observables
corresponding to a charged fluid in absence of RR. On the other hand, the
asymptotic contributions of the RR effect have been isolated on the rhs,
which allows one to interpret them as source terms due to extra forces
acting on the unperturbed fluid. In particular, the presence of the RR acts
like a non-conservative collisional operator, if we interpret it as a sort
of retarded scattering of the fluid (and therefore, of the single particles
at the kinetic level) with itself.

\bigskip

\section{Lagrangian formulation of the fluid equations}

An important issue concerns the treatment of the non-local contributions
appearing in the fluid equations both in the definitions of the fluid fields
and in the source term in the momentum equation. This requires, in
particular, the explicit representation of the self-potential $\overline{A}%
_{\mu }^{(self)}$ and the EM self-force $\overline{F}_{\mu k}^{\left(
self\right) }$ defined respectively in Eqs.(\ref{atot2}) and (\ref{COVARIANT
FORM}). In fact, in the previous sections these non-local contributions have
been written in a parameter-free representation (integral form), so that
they do not depend on the retarded particle velocity. This allowed us to
perform the velocity integrals in a straightforward way, only in terms of
local 4-velocities, in agreement with the formalism adopted for the
Hamiltonian formulation in standard form.

To treat these non-local terms it is first convenient to represent the fluid
moment equations in Lagrangian form, describing the dynamics of fluid
elements along their respective Lagrangian path (LP). By substituting the
definition (\ref{den2}) in Eq.(\ref{continuity}) we obtain the corresponding
Lagrangian form of the continuity equation, given by%
\begin{equation}
\frac{D}{Ds}n+n\partial _{\mu }U^{\mu }=0,  \label{cont-lag}
\end{equation}%
where $\frac{D}{Ds}\equiv U^{\mu }\left( r\left( s\right) \right) \partial
_{\mu }$ is the convective Lagrangian derivative along the LP of the fluid
element parametrized in terms of the arc-length $s$, and $U^{\mu }\left(
r\left( s\right) \right) =\frac{dr^{\mu }\left( s\right) }{ds}$. Similarly,
writing the stress-energy tensor $T^{\mu \nu }\left( r\right) $ as $T^{\mu
\nu }\left( r\right) =nU^{\mu }U^{\nu }+P^{\mu \nu }\left( r\right) $, with $%
P^{\mu \nu }\left( r\right) \equiv T^{\mu \nu }\left( r\right) -nU^{\mu
}U^{\nu }$, the energy-momentum equation (\ref{momentum}) can be represented
in Lagrangian form as follows:%
\begin{equation}
n\frac{D}{Ds}U^{\nu }=nF_{\left( tot\right) }^{\nu \mu }U_{\mu }-\partial
_{\mu }P^{\mu \nu }.  \label{euler-lag}
\end{equation}%
Analogous results can be given for the asymptotic equations (\ref{as1}) and (%
\ref{as2}).

With the introduction of the LPs, the parametrization of the non-local
contributions can be easily reached in terms of the LP arc-length $s$.
Consider, for example, the self-potential $\overline{A}_{\mu }^{(self)}$.
This can be expressed as%
\begin{equation}
\overline{A}_{\mu }^{(self)}\left( r,\left[ r\right] \right) \equiv
2q\int_{1}^{2}ds^{\prime }\frac{dr_{\mu }^{\prime }}{ds^{\prime }}\delta (%
\widetilde{R}^{\mu }\widetilde{R}_{\mu }-\sigma ^{2}),
\end{equation}%
where by definition now $\frac{dr_{\mu }^{\prime }}{ds^{\prime }}=U^{\mu
}\left( r\left( s^{\prime }\right) \right) $ is defined along a fluid
element LP. Then, by expressing the Dirac-delta function as%
\begin{equation}
\delta (\widetilde{R}^{\mu }\widetilde{R}_{\mu }-\sigma ^{2})=\frac{1}{%
\left\vert 2\widetilde{R}^{\alpha }U_{\alpha }\right\vert }\delta \left(
s^{\prime }-s+s_{ret}\right) ,
\end{equation}%
it follows that $\overline{A}_{\mu }^{(self)}$ can be equivalently written
in the integrated form as%
\begin{equation}
\overline{A}_{\mu }^{(self)}\left( r,\left[ r\right] \right) =q\left[ \frac{%
U_{\mu }\left( r\left( s^{\prime }\right) \right) }{\left\vert \widetilde{R}%
^{\alpha }U_{\alpha }\left( r\left( s^{\prime }\right) \right) \right\vert }%
\right] _{s^{\prime }=s-s_{ret}},  \label{self_integral}
\end{equation}%
with $\widetilde{R}^{\alpha }$ being the displacement vector defined along a
LP. In particular, in agreement with the Einstein Causality Principle, the
retarded time $s_{ret}=s-s^{\prime }$ is the positive root of the delay-time
equation%
\begin{equation}
\widetilde{R}^{\mu }\widetilde{R}_{\mu }-\sigma ^{2}=0.  \label{delay}
\end{equation}

An analogous derivation can be carried out also for the self-force $%
\overline{F}_{\mu k}^{\left( self\right) }$, giving the following result%
\begin{equation}
\overline{F}_{\mu k}^{\left( self\right) }\left( r,\left[ r\right] \right)
=-2q\left\{ \frac{1}{\left\vert \widetilde{R}^{\alpha }U_{\alpha }(s^{\prime
})\right\vert }\frac{D}{Ds^{\prime }}X_{\mu k}\left( r\left( s^{\prime
}\right) \right) \right\} _{s^{\prime }=s-s_{ret}},  \label{fself}
\end{equation}%
where%
\begin{equation}
X_{\mu k}\left( r\left( s^{\prime }\right) \right) \equiv \left[ \frac{%
U_{\mu }(r\left( s^{\prime }\right) )\widetilde{R}_{k}-U_{k}(r\left(
s^{\prime }\right) )\widetilde{R}_{\mu }}{\widetilde{R}^{\alpha }U_{\alpha
}(r\left( s^{\prime }\right) )}\right] .
\end{equation}%
Again, this expression must be intended as a parametrization defined along a
fluid element LP.

\bigskip

We conclude by commenting on the following remarkable aspects of the theory
presented here.

1) The fluid equations with the inclusion of the non-local effect related to
the EM RR have been derived in a closed analytical form in both Eulerian and
Lagrangian formulations. In particular, it follows that the fluid dynamics
of the non-local kinetic system is intrinsically non-local too.

2) Non-local contributions of the RR appear both in explicit and implicit
contributions, through the definitions of the fluid fields as velocity
moments of the KDF.

3) From the point of view of the fluid description, it follows that the
natural setting for the treatment of the non-local fluid equations is given
by the Lagrangian formulation and the concept of LPs. This is a consequence
of the fact that the exact moment equations are of delay-type. In fact, in
order to properly deal with the non-local contributions of the RR the
parametrization of the retarded effects in terms of the arc-length of the
corresponding LPs is needed. It follows that the dynamics of a generic fluid
element along its LP is related to the EM RR effect produced at the retarded
time along the LP itself.

\bigskip

\section{Asymptotic approximation}

In the previous sections we derived an exact formulation for both kinetic
and fluid theories describing systems of relativistic charged particles
subject to the EM RR self-interaction. In particular, we have pointed out
that the kinetic and fluid equations are of delay-type, and therefore
intrinsically non-local, due to the characteristic feature of the RR effect
of being a non-local retarded effect. The retarded proper time is determined
by Eq.(\ref{delay}) in agreement with the causality principle. Notice that
this equation has formally the same expression for the single-particle or
the kinetic dynamics and for the fluid equations in Lagrangian form (see
also Paper I). By inspecting Eq.(\ref{delay}) it is easy to realize that the
order of magnitude of the delay-time is approximately $s_{ret}\sim \sigma /c$%
, and therefore very small for classical elementary particles. The smallness
of the retarded time may represent a serious problem for the practical
implementation of the exact theory presented here. In fact, the retarded
time associated to the RR can be orders of magnitude smaller than any other
characteristic time for most of relevant physical situations. The question
is of primary importance, for example, for the actual numerical integration
of the exact fluid equations.

In view of these considerations, in this section we provide asymptotic
estimations of the non-local terms appearing in the moment equations, which
allow one to overcome the difficulty connected with the finite delay-time
intervals carried by the RR phenomenon. This requires to introduce a
suitable asymptotic expansion of the exact non-local terms by means of
approximations in which the self-interaction contributions are all expressed
only through local quantities. The result has potential interest also in
relation to the use of Eulerian integration schemes for the fluid equations
with the inclusion of the RR effect.

Specifically, the present analysis requires to develop an asymptotic
approximation which involves the treatment of the delay-time $s_{ret}$. This
is accomplished within the short delay-time ordering approximation given by
Eq.(\ref{ordering -1}). In the following we shall work adopting the
Lagrangian representation form for the fluid equations. To perform the
asymptotic expansion, we assume that both the external EM field acting on
each fluid element and the macroscopic fluid fields associated to the
kinetic system are smooth function of the coordinate 4-position vector $%
r^{\alpha }$, namely they are of class $C^{k}$, with $k\geq 2$. The result
of the asymptotic approximation for the terms associated to the RR
self-interaction is provided by the following theorem.

\bigskip

\textbf{THM.7 - First-order, short delay-time asymptotic approximation
(present-time expansion).}

\emph{Given validity of the asymptotic ordering (\ref{ordering -1}) and the
smoothness assumptions for the external EM and the fluid fields, neglecting
corrections of order }$\epsilon ^{n},$ \emph{with} $n\geq 1$ \emph{%
(first-order approximation)}$,$\emph{\ it follows that:}

\emph{T7}$_{1}$\emph{) The retarded self-potential }$\overline{A}_{\mu
}^{(self)}$ \emph{defined in Eq.(\ref{self_integral}) can be expanded in a
neighborhood of }$s$\emph{\ as follows:}%
\begin{equation}
\overline{A}_{\mu }^{(self)}=\left. \overline{A}_{\mu }^{(self)}\right\vert
_{s}\left[ 1+O(\epsilon )\right] ,  \label{asin1}
\end{equation}%
\emph{where the present-time leading-order contribution }$\left. \overline{A}%
_{\mu }^{(self)}\right\vert _{s}$ \emph{is given by}%
\begin{equation}
\left. \overline{A}_{\mu }^{(self)}\right\vert _{s}=q\left[ \frac{1}{\sigma }%
U_{\mu }\left( r\left( s\right) \right) -\frac{D}{Ds}U_{\mu }\left( r\left(
s\right) \right) \right] ,
\end{equation}%
\emph{with }$\frac{D}{Ds}$ \emph{being the convective derivative along a
fluid element Lagrangian path.}

\emph{T7}$_{2}$\emph{) Concerning Eq.(\ref{euler-lag}), let us define the
vector field }$K_{\mu }$ \emph{as follows:}%
\begin{equation}
K_{\mu }\equiv \frac{q}{m_{o}c^{2}}\overline{F}_{\mu \nu }^{\left(
self\right) }U^{\nu },  \label{gmu}
\end{equation}%
\emph{with }$\overline{F}_{\mu \nu }^{\left( self\right) }$\emph{\ defined
in Eq.(\ref{fself}). Then, in a neighborhood of }$s$\emph{, }$K_{\mu }$\emph{%
\ can be expanded as follows:}%
\begin{equation}
K_{\mu }=\left. K_{\mu }\right\vert _{s}\left[ 1+O(\epsilon )\right] ,
\label{gs}
\end{equation}%
\emph{where the present-time leading-order contribution }$\left. K_{\mu
}\right\vert _{s}$ \emph{is given by}%
\begin{equation}
\left. K_{\mu }\right\vert _{s}=\left\{ -\frac{1}{\sigma }\frac{q^{2}}{%
m_{o}c^{2}}\frac{D}{Ds}U_{\mu }\left( r\left( s\right) \right) +g_{\mu
}\right\} ,
\end{equation}%
\emph{with }$g_{\mu }$\emph{\ denoting the 4-vector}%
\begin{equation}
g_{\mu }=\frac{2}{3}\frac{q^{2}}{m_{o}c^{2}}\left[ \frac{D^{2}}{Ds^{2}}%
U_{\mu }-U_{\mu }(s)U^{k}(s)\frac{D^{2}}{Ds^{2}}U_{k}\right] .
\end{equation}

\emph{Proof} - The proof of T7$_{1}$) and T7$_{2}$)\emph{\ }can be reached
by introducing a Taylor expansion in terms of the retarded time $s^{\prime }$
for the relevant quantities appearing in Eqs.(\ref{self_integral}) and (\ref%
{fself}). In particular, for the 4-velocity $U_{\mu }\left( r\left(
s^{\prime }\right) \right) $ and the displacement vector $\widetilde{R}^{k}$
we obtain respectively%
\begin{equation}
U_{\mu }\left( r\left( s^{\prime }\right) \right) \cong U_{\mu }\left(
r\left( s\right) \right) -(s-s^{\prime })\frac{D}{Ds}U_{\mu }\left( r\left(
s\right) \right) +\frac{(s-s^{\prime })^{2}}{2}\frac{D^{2}}{Ds^{2}}U_{\mu
}\left( r\left( s\right) \right) +O\left( \epsilon ^{3}\right)
\end{equation}%
and%
\begin{equation}
\widetilde{R}^{k}\cong (s-s^{\prime })U^{k}-\frac{(s-s^{\prime })^{2}}{2}%
\frac{D}{Ds}U^{k}+\frac{(s-s^{\prime })^{3}}{6}\frac{D^{2}}{Ds^{2}}%
U^{k}+O\left( \epsilon ^{4}\right) ,
\end{equation}%
while for the time delay $s-s^{\prime }\equiv s_{ret}$ we get%
\begin{equation}
s-s^{\prime }\cong \sigma +O\left( \epsilon ^{2}\right) .  \label{alfa}
\end{equation}%
By substituting these expansions in Eqs.(\ref{self_integral}) and (\ref%
{fself}), after straightforward calculations the asymptotic solutions (\ref%
{asin1}) and (\ref{gs}) follow identically.

\textbf{Q.E.D.}

\bigskip

We notice that the asymptotic expansion of the self-potential illustrated in
THM.7 is required to reduce the non-local dependencies which are implicit in
the definition of the fluid fields through the KDF. On the other hand,
within the approximation obtained in THM.7 for the 4-vector $K_{\mu }$, the
RR equation (\ref{euler-lag}) reduces to a local third-order ordinary
differential equation. In particular, Eq.(\ref{fself}) in THM.7 represents
the analogue of the LAD equation for the single-particle dynamics, which
contains the first derivative of the particle 4-acceleration (see also
related discussion in Paper I). In view of this similarity, the asymptotic
solution (\ref{gs}) can be further simplified adopting a second
reduction-step of the same kind of that which leads to the LL form of the
self-force for single charged particles \cite{LL}. This is obtained by
assuming that the RR effect is only a small correction to the motion of the
fluid. As a consequence, an iterative approximation can be adopted which
permits to represent the self-force in terms of the instantaneous fluid
forces. The latter include both the external EM field and the pressure
forces. In particular, according to this method, to leading-order for the
fluid 4-acceleration we have%
\begin{equation}
\frac{D}{Ds}U^{\nu }=F_{\left( ext\right) }^{\nu \mu }U_{\mu }-\frac{1}{n}%
\partial _{\mu }P^{\mu \nu },
\end{equation}%
where, for brevity we have introduced the notation%
\begin{equation}
F_{\left( ext\right) }^{\nu \mu }\equiv \frac{q}{m_{o}c^{2}}\overline{F}%
^{\left( ext\right) \nu \mu }.
\end{equation}%
The iteration gives%
\begin{eqnarray}
\frac{D^{2}}{Ds^{2}}U^{\nu } &=&\partial _{l}F_{\left( ext\right) }^{\nu \mu
}U_{\mu }U^{l}+F_{\left( ext\right) }^{\nu \mu }\left( F_{\left( ext\right)
\mu l}U^{l}-\frac{1}{n}\partial _{l}P_{\mu }^{l}\right) +  \notag \\
&&+\frac{1}{n}\partial _{\mu }P^{\mu \nu }U^{l}\partial _{l}\ln n-\frac{1}{n}%
U^{l}\partial _{l}\partial _{\mu }P^{\mu \nu }.  \label{LLfluid}
\end{eqnarray}%
Substituting this expansion in Eq.(\ref{gs}) and invoking the symmetry
property of the Faraday tensor provides for the first-order term $\left.
K_{\mu }\right\vert _{s}$ the following approximation:%
\begin{equation}
\left. K_{\mu }\right\vert _{s}\simeq \frac{q^{2}}{m_{o}c^{2}}\left\{ -\frac{%
1}{\sigma }\left[ \frac{q}{m_{o}c^{2}}\overline{F}_{\mu \nu }^{\left(
ext\right) }U^{\nu }-\frac{1}{n}\partial _{\nu }P_{\mu }^{\nu }\right] +%
\frac{2q}{3m_{o}c^{2}}h_{\mu }^{\left( 1\right) }+\frac{2}{3}h_{\mu
}^{\left( 2\right) }\right\} ,  \label{kll}
\end{equation}%
where the first term on the rhs represents the mass-renormalization
contribution, and $h_{\mu }^{\left( 1\right) }$\ denotes the 4-vector%
\begin{equation}
h_{\mu }^{\left( 1\right) }=\partial _{l}\overline{F}_{\mu \nu }^{\left(
ext\right) }U^{\nu }U^{l}-\frac{q}{m_{o}c^{2}}\overline{F}_{\mu \nu
}^{\left( ext\right) }\overline{F}^{\left( ext\right) \nu l}U_{l}+\frac{q}{%
m_{o}c^{2}}\left( \overline{F}_{kl}^{\left( ext\right) }U^{l}\right) \left(
\overline{F}^{\left( ext\right) k\nu }U_{\nu }\right) U_{\mu },  \label{h11}
\end{equation}%
while $h_{\mu }^{\left( 2\right) }$ is given by%
\begin{eqnarray}
h_{\mu }^{\left( 2\right) } &=&-\frac{q}{m_{o}c^{2}}\frac{1}{n}\overline{F}%
_{\mu \beta }^{\left( ext\right) }\partial _{l}P^{l\beta }+\frac{1}{n}%
\partial _{\nu }P_{\mu }^{\nu }U^{l}\partial _{l}\ln n-\frac{1}{n}%
U^{l}\partial _{l}\partial _{\nu }P_{\mu }^{\nu }+  \label{h22} \\
&&+\frac{q}{m_{o}c^{2}}\frac{1}{n}U_{\mu }U^{k}\overline{F}_{k\beta
}^{\left( ext\right) }\partial _{l}P^{l\beta }-\frac{1}{n}U_{\mu
}U^{k}\partial _{\nu }P_{k}^{\nu }U^{l}\partial _{l}\ln n+\frac{1}{n}U_{\mu
}U^{k}U^{l}\partial _{l}\partial _{\nu }P_{k}^{\nu }.  \notag
\end{eqnarray}%
Eq.(\ref{kll}) represents the fluid analogue of the LL approximation of the
self-force holding for single particle dynamics, with the
mass-renormalization term retained. In particular here we notice that:

1) Eq.(\ref{kll}) provides a local approximation of the fluid self-force
carrying the contribution of the RR effect. In contrast to Eq.(\ref{gs}),
thanks to the iterative reduction procedure only second-order derivatives of
the position vector appear in this approximation.

2) For consistency, Eq.(\ref{kll}) must be evaluated adopting the asymptotic
expansion (\ref{asin1}) also for the evaluation of the self-potential
entering the definition of the fluid fields through the canonical momenta $%
P_{\mu }$ in the KDF.

3) Moreover, consistent with the approximation in which the RR
self-potential is small with respect to the external EM potential, also the
asymptotic approximation (\ref{series1}) can be adopted, which allows one to
treat explicitly in an asymptotic way all the implicit RR contributions.

4) Finally, collecting together the analytical approximations provided by
Eqs.(\ref{series1}), (\ref{asin1}) and (\ref{kll}), the fluid equations are
reduced to a set of asymptotic local second-order PDEs. This provides a
convenient representation also for Eulerian implementation schemes of the
same equations.

The detail comparison of Eqs.(\ref{LLfluid})-(\ref{h22}) with the literature
is discussed in the next section.

\bigskip

\subsection{Retarded-time asymptotic expansion}

Despite the previous considerations, it is worth pointing out that, formally
also for the fluid equations, an analogous result to THM.7 can be given.
This is based on performing a Taylor expansion of the fluid RR force based
on the retarded-time approximation. In this case, it is found that Eq.(\ref%
{self_integral}) is approximated as%
\begin{equation}
\overline{A}_{\mu }^{(self)}=\left. \overline{A}_{\mu }^{(self)}\right\vert
_{s^{\prime }}\left[ 1+O(\epsilon )\right] ,  \label{ret1}
\end{equation}%
where the retarded-time leading-order contribution $\left. \overline{A}_{\mu
}^{(self)}\right\vert _{s^{\prime }}$ is simply given by%
\begin{equation}
\left. \overline{A}_{\mu }^{(self)}\right\vert _{s^{\prime }}=\frac{q}{%
\sigma }U_{\mu }\left( r\left( s^{\prime }\right) \right) ,
\end{equation}%
while Eq.(\ref{fself}) for the self-force, written in terms of $K_{\mu }$
defined in Eq.(\ref{gmu}), becomes%
\begin{equation}
K_{\mu }=\left. K_{\mu }\right\vert _{s^{\prime }}\left[ 1+O(\epsilon )%
\right] ,
\end{equation}%
where the retarded-time leading-order contribution $\left. K_{\mu
}\right\vert _{s^{\prime }}$ is now given by%
\begin{equation}
\left. K_{\mu }\right\vert _{s}=\left\{ -\frac{q^{2}}{\sigma m_{o}c^{2}}%
\frac{D}{Ds^{\prime }}U_{\mu }\left( r\left( s^{\prime }\right) \right)
+g_{\mu }^{\prime }\left( r\left( s^{\prime }\right) \right) \right\} ,
\end{equation}%
with $g_{\mu }^{\prime }$\ denoting here the 4-vector%
\begin{equation}
g_{\mu }=-\frac{1}{3}\frac{q^{2}}{m_{o}c^{2}}\left[ \frac{D^{2}}{Ds^{\prime
2}}U_{\mu }\left( r\left( s^{\prime }\right) \right) -U_{\mu }\left( r\left(
s^{\prime }\right) \right) U^{k}\left( r\left( s^{\prime }\right) \right)
\frac{D^{2}}{Ds^{\prime 2}}U_{k}\left( r\left( s^{\prime }\right) \right) %
\right] .  \label{ret2}
\end{equation}%
This alternative expansion has the distinctive advantage (with respect to
the present-time expansion) of retaining all the physical properties of the
exact fluid equations for the treatment of RR delay-time effects. This
alternative formulation is relevant for comparisons with the point-particle
treatment.

\section{Discussion and comparisons with literature}

In this section we analyze in detail the physical properties of the kinetic
and fluid theory developed for the EM RR problem, providing also a
comparison with the literature. This concerns, in particular, the recent
paper by Berezhiani et al. \cite{Ma2004}, where an analogous research
program is presented for the relativistic hydrodynamics with RR based on the
LL solution of the self-force.

\subsection{Kinetic theory}

Let us start by considering the kinetic theory. The solution here obtained
has the following key features:

1) The kinetic theory adopts the Hamiltonian formulation of the RR problem
here developed. The result is based on the exact analytical solution for the
EM self-potential of finite-size charged particles, obtained in Paper I and
Appendix A.

2) The kinetic theory is here developed for systems of charged particles
subject to an external mean-field EM interaction and the RR self-interaction
produced by the same particles. Due to the non-local property of the RR
interaction, the formulation of kinetic theory is non-trivial. For this
purpose, in contrast to previous literature, an axiomatic formulation of CSM
is adopted. Its key element is the introduction of a suitable definition for
the Lorentz-invariant probability-measure in the particle extended
phase-space. As a consequence, the corresponding Liouville-Vlasov kinetic
equation with the inclusion of the exact RR effect is achieved in
Hamiltonian form, namely in such a way to preserve the phase-space canonical
measure. For comparison, instead, previous literature approaches dealt with
measure non-preserving phase-space dynamics.

3) In particular, the kinetic theory has been developed within the canonical
formalism representing the KDF in terms of the canonical state $\mathbf{y}%
\equiv \left( r^{\mu },P_{\mu }\right) $. For reference, in Appendix B the
connection with the corresponding non-canonical treatment is provided. This
in turn implies that non-local contributions associated to the
self-potential (\ref{atot2}) enter implicitly in the definition of the
corresponding fluid moments (\ref{den1})-(\ref{den3}). This is made possible
only within the framework of the present exact formulation, in which the
analytical solution for the self-potential is by construction non-divergent.
This feature departs from recent approaches where instead non-Hamiltonian
formulations were adopted, based on the LL point-like approximation of the
RR self-force. In such a case in fact, the explicit dependence of the KDF in
terms of the EM self-potential cannot be retained.

4) Both the RR equation for single-particle dynamics and the kinetic
equation for the KDF are of delay-type, reflecting the characteristic nature
of the RR phenomenon. This property is completely missing from the previous
literature on the subject, exclusively based on the LL local asymptotic
approximation.

\bigskip

\subsection{Fluid theory}

For what concerns the fluid treatment, we notice that:

1) Both the fluid fields and the fluid moment equations retain the standard
form (available in the absence of RR effects) and can be equivalently
represented in Eulerian or Lagrangian form. This follows from the exact
representation here adopted both for the RR self-potential and the RR
self-force. In both cases the only non-local dependencies are those
associated to the position 4-vector.

2) The exact fluid equations with the inclusion of the RR effect are
delay-type PDEs. Because of this feature, their natural representation
appears to be the Lagrangian form. In fact, the integration along the LPs
must be in principle performed taking into account the retarded RR
interaction.

3) From the exact theory presented here it follows that each fluid equation
of a given order does not depend on fluid fields of higher orders. For
example, the momentum equation contains only second-order tensor fields,
identified respectively with the plasma stress-energy tensor and the EM
Faraday tensor. This result contrasts with the treatment given in Ref.\cite%
{Ma2004} where instead the asymptotic formulation based on the LL equation
leads to moment equations involving higher-order tensor fields (for
comparison, see also the related discussion in Appendix B).

4) If a kinetic closure is chosen, then the fluid moments appearing in the
fluid equations are all uniquely determined. In particular, the
stress-energy tensor is prescribed in terms of the KDF. This implies that
both implicit and explicit contributions of the RR effect appear in the
resulting equations, carried respectively by the fluid fields and the EM
self-force in the momentum equation. Remarkably, kinetic closure is achieved
prescribing solely the pressure contribution carried by the stress-energy
tensor. Instead, in the approach of Ref.\cite{Ma2004} the closure conditions
involve generally also the specification of higher-order moments of the KDF.

5) An important feature of the exact fluid equations here obtained is that
they can in principle be exactly implemented numerically adopting a
Lagrangian scheme.

6) A remarkable aspect of the present theory is that the relevant asymptotic
expansions are performed only \textquotedblleft a
posteriori\textquotedblright\ after integration over the velocity space.
This means that the approximations involved are introduced only on the
configuration space-variables (i.e., the fluid fields) and not on the
phase-space KDF. In particular, a convenient approximation is the one
obtained in the short delay-time ordering, which reduces the non-local
dependencies to local terms. As a consequence, the introduction of
higher-order moments is ruled out by construction.

\bigskip

\subsection{Comparison with point-particle treatments}

The relevant comparison here is represented by Ref.\cite{Ma2004}. Such an
approach is based on the adoption of the LL equation for the single-particle
dynamics for the construction of the relativistic Vlasov-Maxwell
description. The corresponding moment equations can be in principle adopted
for the construction of a closed set of fluid equations. This requires
however the specification of suitable closure-conditions. Let us briefly
point out the novel features of the current treatment for what concerns the
adoption of the finite-size particle model in the construction of the
kinetic and fluid descriptions. In detail:

1) Both in the kinetic and fluid treatments the RR force is taken into
account by means of a non-local interaction. This is an intrinsic feature of
the assumed finite extension of the charged particle. In the fluid
treatment, in particular, as shown above, the RR force can be parametrized
in terms of the past Lagrangian fluid velocity and position. This permits to
treat consistently the causal delay-time effects due to the finite-size of
the particles.

2) In validity of the asymptotic ordering given by Eq.(\ref{ordering -1}),
an asymptotic retarded-time Hamiltonian approximation of the RR force based
on a retarded-time expansion has been given for the fluid equations. This
approximation preserves the basic physical features of the solution based on
the exact form of the RR self-force.

3) If the present-time asymptotic expansion is performed on the exact fluid
moment equations, the resulting expression of the fluid RR force obtained
adopting the finite-size charge model appears different from that given in
Ref.\cite{Ma2004}.

These conclusions enable us to carry out a detailed comparison with the
literature, emphasizing the basic differences between kinetic and fluid
treatments based on finite-size and point particles.

A) Kinetic theory.

The kinetic equation adopted in Ref.\cite{Ma2004} is based on the LL
equation (see therein Eqs.7 and 8). This means that the RR force in this
approximation is non-conservative, non-variational and therefore
non-Hamiltonian. In addition the LL equation: 1) does not retain finite
delay-time effects characteristic of the RR phenomenon; 2) is not valid in
the case of strong EM fields, where the iterative reduction scheme on which
it is based, may fail; 3) ignores mass-renormalization effects (which are
incompatible with the point-particle model). In contrast, the treatment of
the relativistic Vlasov kinetic equation obtained here (see the Eulerian and
Lagrangian equations (\ref{euler-kinm}) and (\ref{lagr-kinm})) is
qualitatively different. In fact, even if the resulting RR equation remains
a second-order ODE, it is conservative, variational, Hamiltonian and applies
for arbitrary external EM fields. Further remarkable aspects are related to
the adoption of the finite-size charge model, in which the charge and mass
distributions have the same support. As a consequence, in this case the self
4-potential is everywhere well-defined, contrary to the point particle
model. In addition, this is prescribed analytically (see Appendix A), a
feature which allows one to treat consistently the RR delay-time effects.

B) Fluid theory.

The fluid treatment here obtained is provided by the Eulerian Eqs.(\ref%
{continuity})-(\ref{momentum}) or the equivalent Lagrangian equations (\ref%
{cont-lag})-(\ref{euler-lag}). The latter, considered as fluid equations,
are manifestly not closed. However, the Hamiltonian formulation achieved
here and holding for finite-size particles allows one to achieve a
physically consistent kinetic closure condition, by prescribing uniquely the
pressure tensor $P_{\mu \nu }$ in Eq.(\ref{euler-lag}). We stress that in
our treatment no higher-order moments need to be specified. In contrast, the
corresponding Euler equation reported in Ref.\cite{Ma2004} (see Eqs.11 and
12 therein) actually depends also on a third-order tensor moment, which must
be prescribed (see comments in Sec.IIIA of Ref.\cite{Ma2004}). Let us now
consider the asymptotic fluid treatments based on the present theory. These
can be achieved invoking either the present-time or the retarded-time
asymptotic expansions (see Section X). The first expansion is mostly
relevant for comparisons with Ref.\cite{Ma2004} (given in THM.7) and enables
one to achieve a local approximation of the delay-time effects carried by
the RR force. However, remarkably, the resulting asymptotic fluid equations (%
\ref{LLfluid})-(\ref{h22}) remain qualitatively different from the
corresponding ones given in Ref.\cite{Ma2004}. In particular: 1) no
higher-order moments appear after performing the Taylor expansion and the
iteration scheme discussed after THM.7; 2) a non-vanishing mass-correction
contribution is now included (see first term on the rhs of Eq.(\ref{kll})).
Finally, we mention that the retarded-time asymptotic expansion given by
Eqs.(\ref{ret1})-(\ref{ret2}) provides a novel approximation which retains
basic properties of the exact solution. In particular: 1) it only applies
for finite-size particles; 2) it relies on the Hamiltonian formulation of
the RR problem and of the Vlasov-Maxwell treatment; 3) it permits to retain
transient-time and delay-time effects; 4) it takes into account retarded
mass-correction effects; 5) in this approximation the natural fluid
description is Lagrangian.

\bigskip

\bigskip

\section{Conclusions}

In this paper, novel results have been obtained concerning the kinetic and
fluid descriptions of relativistic collisionless plasmas with the inclusion
of EM RR effects.

Relevant consequences of the variational form of the EM RR equation
previously achieved for classical finite-size charged particles have been
investigated. It has been shown that the non-local RR problem admits both
Lagrangian and Hamiltonian representations in standard form, defined
respectively in terms of effective Lagrangian and Hamiltonian functions. A
remarkable novel feature of the theory concerns the development of a
Hamiltonian retarded-time expansion of the RR force, which applies in
validity of the short delay-time asymptotic ordering. On such a basis, the
axiomatic formulation of classical statistical mechanics for relativistic
collisionless plasmas with the inclusion of non-local RR effects has been
presented. As a major result, the kinetic theory for such a system has been
formulated in standard Hamiltonian form. The Liouville-Vlasov equation has
been proved to hold in the extended phase-space, subject to non-local RR
self-interactions. Remarkably, the non-local effects have been proved to
enter the relativistic kinetic equation only through the retarded particle
4-position. As a consequence, the corresponding fluid moment equations can
be determined in standard way by integration over the space of canonical
momenta and cast both in Eulerian and Lagrangian forms. It has been pointed
out that the exact relativistic fluid equations are intrinsically of
delay-type and contain both implicit and explicit non-local contributions
associated to the RR effect. The issue concerning the problem of fluid
closure conditions has been discussed. In contrast with previous literature,
it is found that in the present approach the closure conditions remain the
standard ones, i.e., as in the absence of RR effects. Hence, the
specification of higher-order moments of the KDF, for a given moment
equation, is not required. Finally, appropriate approximations have been
obtained for the fluid equations by employing \textquotedblleft a
posteriori\textquotedblright\ the relevant asymptotic expansions applicable
in the short delay-time ordering. This allows one to reduce the exact
non-local equations either to a set of local PDEs or to retarded PDEs still
retaining finite delay-time effects.

The theory here developed has potential wide-ranging applications which
concern the study of relativistic astrophysical plasmas for which RR
emission processes are important. This involves, for example, plasmas in
accretion disks, relativistic jets and active galactic nuclei. Other
possible applications are also suggested for the case of laboratory plasmas
generated in the presence of pulsed-laser sources.

\bigskip

\section{Appendix A: Integral representation for $A_{\protect\mu }^{(self)}$
- Case of a non-rotating spherical-shell charged particle}

In this Appendix we determine explicitly the integral representation of $%
A_{\mu }^{(self)}$ for a non-rotating finite-size charged particle described
by the model introduced in Paper I. We first remark that Eqs.(\ref{eee1a})
and (\ref{eee2a}) can be written for a spherically-symmetric charged
particle of radius $\sigma >0$ as%
\begin{eqnarray}
\xi ^{\mu }\xi _{\mu } &=&-\sigma ^{2},  \label{eee1} \\
\xi _{\mu }u^{\mu }(s) &=&0.  \label{eee2}
\end{eqnarray}%
Eq.(\ref{eee1}) defines the boundary $\partial \Omega $ on which the charge
and mass are uniformly distributed, while Eq.(\ref{eee2}) represents the
constraint of rigidity for the finite-size particle. We can use the
information from Eq.(\ref{eee1}) to define the \emph{internal} and the \emph{%
external} domains with respect to the mass and charge distributions. In
particular, in terms of the generic displacement 4-vector $X^{\mu }\in M^{4}$
defined as
\begin{equation}
X^{\mu }=r^{\mu }-r^{\mu }\left( s\right)
\end{equation}%
and subject to the constraint%
\begin{equation}
X^{\mu }u_{\mu }(s)=0,
\end{equation}%
the following relations hold:%
\begin{eqnarray}
X^{\mu }X_{\mu } &\leq &-\sigma ^{2}\emph{\ :external}\text{ }\emph{domain,}
\label{ext} \\
X^{\mu }X_{\mu } &>&-\sigma ^{2}\emph{\ :internal}\text{ }\emph{domain,}
\notag \\
X^{\mu }X_{\mu } &=&\xi ^{\mu }\xi _{\mu }=-\sigma ^{2}\emph{\ :boundary}%
\text{ }\emph{location.}  \notag
\end{eqnarray}%
As proved in Ref.\cite{Cremaschini2011}, for the evaluation of the action
integral $S_{C}^{(self)}$ it is sufficient to know the solution of $A_{\mu
}^{(self)}$ in the external domain. In this domain the EM self 4-potential
generated by the non-rotating finite-size particle must necessarily coincide
with that of a point particle carrying the same total mass and charge. The
retarded 4-potential of a point charge represents a well-known result in the
literature \cite{Jak}. This can be easily obtained by means of the Green
function approach. In particular, introducing the retarded Green function of
a point particle $G(r-r^{\prime })$, the self-potential $A_{\mu }^{(self)}$
takes the form%
\begin{equation}
A^{(self)\mu }(r)=\frac{4\pi }{c}\int d^{4}r^{\prime }G(r-r^{\prime })j^{\mu
}(r^{\prime }),  \label{D-1bis}
\end{equation}%
where $G(r-r^{\prime })$ is symmetric in both $r$ and $r^{\prime },$ is
non-vanishing only for $r^{0}<r^{\prime 0}$ and satisfies in this domain the
boundary-value problem%
\begin{equation}
\left\{
\begin{array}{c}
\square G(r-r^{\prime })=\delta ^{(4)}\left( r-r^{\prime }\right) , \\
G(r-r^{\prime })=0.%
\end{array}%
\right.
\end{equation}%
As a consequence, written in integral form, the self-potential becomes%
\begin{equation}
A^{(self)\mu }(r,q)=2q\int_{1}^{2}dr^{\mu }\delta (\widehat{R}^{\alpha }%
\widehat{R}_{\alpha }),  \label{d2}
\end{equation}%
with%
\begin{equation}
\emph{\ }\widehat{R}^{\alpha }\equiv r^{\alpha }-r^{\alpha }(s).
\end{equation}%
This solution, derived for a point charged particle, also holds for the
rigid finite-size non-rotating spherical shell in the external domain
defined in Eq.(\ref{ext}). As can be seen, this coincides with the result in
Ref.\cite{Cremaschini2011}, where a complete covariant solution for the EM
self 4-potential $A_{\mu }^{(self)}$ holding in both internal and external
domains has been obtained by adopting a derivation based on the principle of
relativity and analogous to that outlined in Ref.\cite{LL} for the point
charge case. Notice that, contrary to the case of point particle, the
self-potential (\ref{d2}) is \textit{well-defined} also on the support of
the charge, namely the ensemble on which the charge is distributed.

\section{Appendix B:\ non-canonical representation}

In this appendix we present the equivalent representation of the kinetic
theory developed in section adopting non-canonical variables. For
definiteness, let us introduce an arbitrary non-canonical phase-space
diffeomorphism from $\Gamma $ to $\Gamma _{\mathbf{w}}$, with $\Gamma _{%
\mathbf{w}}$ denoting a transformed phase-space having the same dimension of
$\Gamma $,%
\begin{equation}
\mathbf{y}\equiv \left( r^{\mu },P_{\mu }\right) \rightarrow \mathbf{w\equiv
w}\left( \mathbf{y}\right) ,  \label{nctra}
\end{equation}%
where, for example, $\mathbf{w}$ can be identified with the non-canonical
state $\mathbf{y}_{nc}\equiv \left( r^{\mu },p_{\mu }\right) $ defined in
Eq.(\ref{series1}) or with $\mathbf{y}_{u}\equiv \left( r^{\mu },u_{\mu
}\right) $. In the second case the transformation, following from Eq.(\ref%
{pp}), is realized by%
\begin{eqnarray}
r^{\mu } &=&r^{\mu },  \label{a1} \\
u_{\mu } &=&P_{\mu }-\frac{q}{c}\left[ \overline{A}_{\mu }^{(ext)}+2%
\overline{A}_{\mu }^{(self)}\right] .  \label{a2}
\end{eqnarray}%
The transformed RR equation in the variables $\mathbf{y}_{u}$ becomes
therefore:%
\begin{eqnarray}
\frac{dr^{\mu }}{ds} &=&u^{\mu },  \label{aa1} \\
\frac{du_{\mu }}{ds} &=&F_{\mu },  \label{aa2}
\end{eqnarray}%
where $F_{\mu }=\frac{\partial p_{\mu }}{\partial r^{\nu }}u^{\nu }-\frac{%
\partial u_{\mu }}{\partial P_{\nu }}\frac{\partial H_{eff}}{\partial r^{\nu
}}$. Denoting now by%
\begin{equation}
f_{1}\left( \mathbf{w}\left( s\right) \right) =\left\vert \frac{\partial
\mathbf{y}\left( s\right) }{\partial \mathbf{w}\left( s\right) }\right\vert
f\left( \mathbf{y}\left( \mathbf{w}\left( s\right) \right) \right)
\label{app0}
\end{equation}%
the KDF mapped onto the transformed phase-space $\Gamma _{\mathbf{w}}$ by
the KDF $f\left( \mathbf{y}\left( s\right) \right) $, the differential
Liouville-Vlasov equation (\ref{liouv}) requires%
\begin{equation}
\frac{d}{ds}\left[ \left\vert \frac{\partial \mathbf{w}\left( s\right) }{%
\partial \mathbf{w}_{0}}\right\vert f_{1}\left( \mathbf{w}\left( s\right)
\right) \right] =0,  \label{app3}
\end{equation}%
where $\mathbf{w}_{0}\equiv \mathbf{w}\left( s_{0}\right) $. At the same
time, Eq.(\ref{liouv}) also implies, thanks to the chain rule:%
\begin{equation}
\frac{d}{ds}f\left( \mathbf{y}\left( \mathbf{w}\left( s\right) \right)
\right) =0,
\end{equation}%
which for consistency delivers the well-known differential identity%
\begin{equation}
\frac{d}{ds}\left[ \left\vert \frac{\partial \mathbf{y}\left( s\right) }{%
\partial \mathbf{w}\left( s\right) }\right\vert \left\vert \frac{\partial
\mathbf{w}\left( s\right) }{\partial \mathbf{w}_{0}}\right\vert \right] =0.
\label{app3bis}
\end{equation}%
From Eq.(\ref{app3}) it follows%
\begin{equation}
\frac{d}{ds}f_{1}\left( \mathbf{w}\left( s\right) \right) +f_{1}\left(
\mathbf{w}\left( s\right) \right) \frac{d}{ds}\ln \left( \left\vert \frac{%
\partial \mathbf{w}\left( s\right) }{\partial \mathbf{w}_{0}}\right\vert
\right) =0.  \label{app4}
\end{equation}%
This equation can be represented, for example, in terms of $\mathbf{w}\equiv
\mathbf{y}_{u}$. In this case, due to the chain rule%
\begin{equation}
\frac{d}{ds}f_{1}\left( \mathbf{w}\left( s\right) \right) =u^{\mu }\frac{%
\partial f_{1}\left( \mathbf{y}_{u}\right) }{\partial r^{\mu }}+F_{\mu }%
\frac{\partial f_{1}\left( \mathbf{y}_{u}\right) }{\partial u_{\mu }},
\end{equation}%
while, thanks to Liouville theorem%
\begin{equation}
\frac{d}{ds}\ln \left( \left\vert \frac{\partial \mathbf{w}\left( s\right) }{%
\partial \mathbf{w}_{0}}\right\vert \right) =\frac{\partial F_{\mu }}{%
\partial u_{\mu }}.
\end{equation}%
As an application of the result, it follows that, if the LL approximation is
introduced for the 4-vector $F_{\mu }$, namely Eqs.(\ref{aa1}) and (\ref{aa2}%
) are replaced with asymptotic equations of the form%
\begin{eqnarray}
\frac{dr_{LL}^{\mu }}{ds} &=&u_{LL}^{\mu },  \label{ab1} \\
\frac{du_{LL}^{\mu }}{ds} &=&F_{LL}^{\mu },  \label{ab2}
\end{eqnarray}%
where $F_{LL}^{\mu }$ is the total EM force in this approximation, then Eq.(%
\ref{app4}) recovers the expression reported in Ref.\cite{Ma2004}. This
provides the connection with the exact canonical theory here developed. We
remark, however, that since the LL equation is only asymptotic, the mapping
between the canonical state $\mathbf{y}\equiv \left( r^{\mu },P_{\mu
}\right) $ and $\mathbf{y}_{LL}\equiv \left( r_{LL}^{\mu },u_{LL\mu }\right)
$ is also intrinsically asymptotic. Therefore, Eqs.(\ref{ab1}) and (\ref{ab2}%
) remain necessarily non-variational and non-canonical.

\end{document}